\documentclass[aps,pra,amssymb,superscriptaddress,twocolumn]{revtex4-1}
\usepackage[pdftex]{graphicx}
\usepackage[dvipsnames]{xcolor}
\usepackage{bm}
\usepackage{amsmath}
\usepackage{ascmac}
\usepackage{color}
\usepackage{setspace}
\usepackage{amssymb}
\usepackage{latexsym}
\usepackage{mathtools}
\usepackage{lipsum}
\usepackage{braket}
\usepackage[colorlinks,linkcolor=RoyalBlue,citecolor=RoyalBlue,urlcolor=RoyalBlue]{hyperref}
\usepackage{MnSymbol}

\begin{document}

\title{Exact Solution of Bipartite Fluctuations in One-Dimensional Fermions} 

\author{Kazuya Fujimoto}
\affiliation{Department of Physics, Institute of Science Tokyo, 2-12-1 Ookayama, Meguro-ku, Tokyo 152-8551, Japan}

\author{Tomohiro Sasamoto}
\affiliation{Department of Physics, Institute of Science Tokyo, 2-12-1 Ookayama, Meguro-ku, Tokyo 152-8551, Japan}

\date{\today}

\begin{abstract}
Emergence of hydrodynamics in quantum many-body systems has recently garnered growing interest. The recent experiment of ultracold atoms [J. F. Wienand {\it et al.},  Nat. Phys.
(2024), doi:10.1038/s41567-024-02611-z] studied emergent hydrodynamics in hard-core bosons using a bipartite fluctuation, which quantifies how the particle number fluctuates in a subsystem. In this Letter, we theoretically study the variance of a bipartite fluctuation in one-dimensional noninteracting fermionic dynamics starting from an alternating state, deriving the exact solution of the variance and its asymptotic linear growth law for the long-time dynamics. To compare the theoretical prediction with the experiment, we generalize our exact solution by incorporating the incompleteness of the initial alternating state, deriving the general linear growth law analytically. We find that it shows good agreement with the experimentally observed variance growth without any fitting parameters. Furthermore, we estimate a time scale for the local equilibration using our exact solution, finding that the time scale is independent of the initial incompleteness. To investigate the interaction effect, we implement numerical studies for the variance growth in interacting fermions, which has yet to be explored experimentally. As a result, we find that the presence of interactions breaks the linear variance growth derived in the noninteracting fermions. Our exact solutions and numerical findings here lay a foundation for growing bipartite fluctuations in quantum many-body dynamics.
\end{abstract}

\maketitle 
{\it Introduction.---}
Relaxation to an equilibrium state is ubiquitous in quantum many-body dynamics, bringing up fundamental and intriguing questions, e.g., how an isolated quantum system relaxes to a thermal equilibrium state \cite{Polkovnikov2011, Eisert2015, Nandkishore2015, Torres2015, Luca2016, Mori2018, Abanin2019,Zongping2022}. Over decades, such quantum thermalization has been intensively studied from broad perspectives, such as the eigenstate thermalization hypothesis \cite{Deutsch1991, Srednicki1994}, integrability~\cite{Kinoshita2006,Gring2012,Cazalilla2006, Rigol2006, Rigol2007, Calabrese2011, Fagotti2013, Langen2015, Ilievski2015,Yijun2018}, many-body localization~\cite{Basko2006, Marko2008, Pal2010, Bardarson2012, Serbyn2013, Huse2014, Schreiber2015, Potter2015, Vasseur2016, Choi2016, Henrik2017,Thomas2019}, and quantum scars~\cite{Bernien2017, Shiraishi2017, Turner2018, Turner2018B, Lin2019, Shibata2020}. To date, many experiments involving ultracold atoms and trapped ions have observed various phenomena related to quantum thermalization \cite{Kinoshita2006, Gring2012, Trotzky2012, Langen2013, Schreiber2015, Langen2015, Govinda2016,  Kaufman2016, Neill2016, Henrik2017, Bernien2017, Yijun2018, Thomas2019}. 

\begin{figure}[b]
\begin{center}
\includegraphics[keepaspectratio, width=8.7cm]{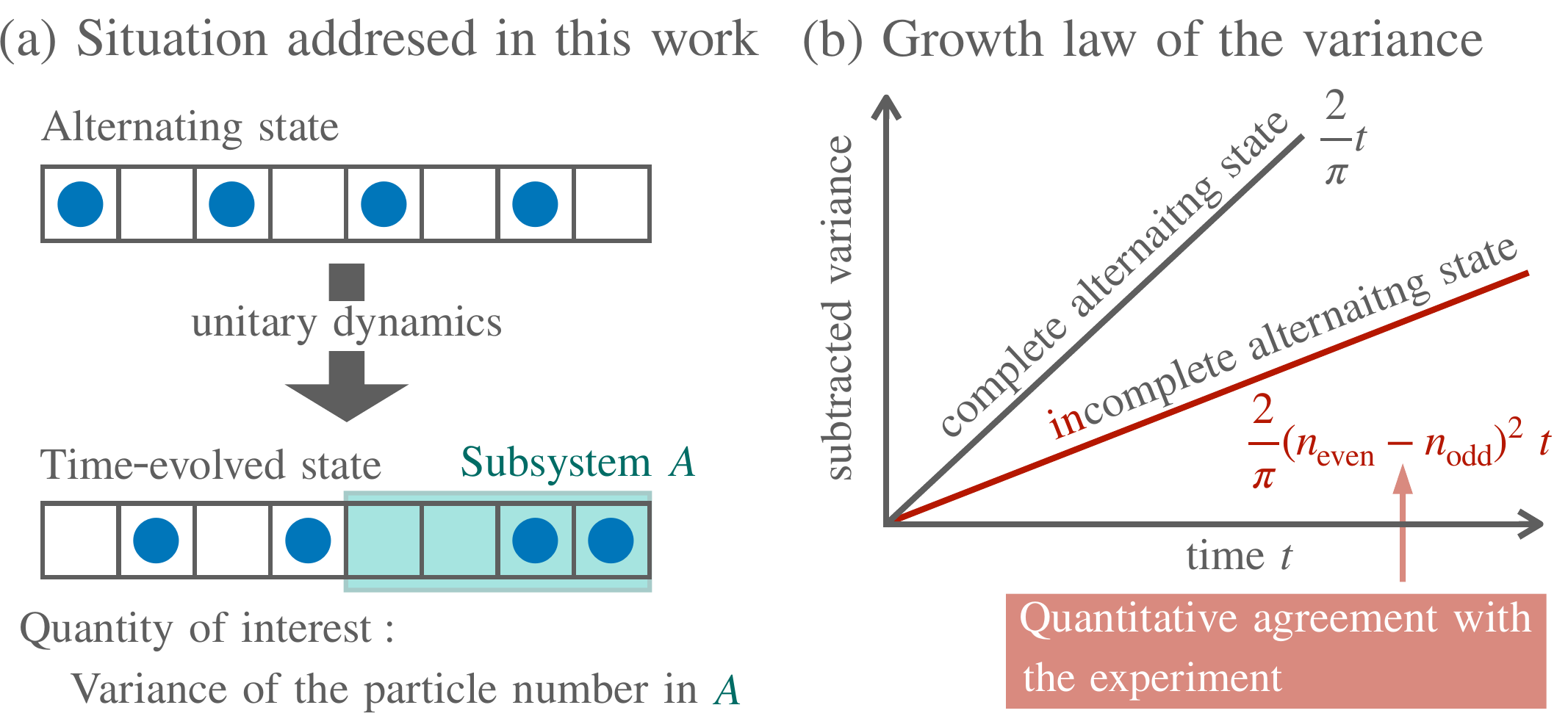}
\caption{
Schematic illustration of the main result. (a) One-dimensional fermions addressed in this work. The blue circles represent fermions, and they occupy every other site at the initial time. This state is referred to as an alternating state. The fermions hop to neighboring sites in time. The quantity of our interest is the variance of the particle number in a subsystem $A$ for a time-evolved state. (b) Schematic for the variance growth. The abscissa and ordinate, respectively, denote time $t$ and a subtracted variance, which is defined by the variance from which one subtracts its initial value. The two solid lines represent our analytical results, namely $2t/\pi$ and $2(n_{\rm even}-n_{\rm odd})^2t/\pi$, obtained by considering the variances growing from the complete and incomplete alternating states. Here, $n_{\rm even}$ and $n_{\rm odd}$ are averaged initial particle numbers at even and odd sites respectively, quantifying the incompleteness of the initial alternating state. 
} 
\label{fig0} 
\end{center}
\end{figure}

Recently, beyond the conventional quantum thermalization, hydrodynamic description based on local equilibrium has rapidly developed in quantum many-body systems. 
This situation is exemplified by the recent observation of electron fluids in strongly interacting systems~\cite{Moll2016, Bandurin2016, Gooth2018, Kumar2017, Sulpizio2019, Steinberg2022} and the success of generalized hydrodynamics being valid for integrable quantum models~\cite{Olalla2016, Bertini2016, Bulchandani2017, Bulchandani18, Doyon2017, Doyon2017_2, Doyon2018, Collura2018, Jacopo2018, Schemmer2019, Sarang2019, Doyon2020_rev, Alba2021, Malvania2021, Bouchoule2022, Essler2023}. When considering such hydrodynamic description, it is of essence to scrutinize how a system approaches a local equilibrium state.
One of the most useful quantities for diagnosing local equilibrium is a bipartite fluctuation that quantifies how a physical quantity in a subsystem fluctuates.
In a recent experiment of ultracold atoms for hard-core bosons~\cite{hydro}, J. F. Wienand {\it et al.} utilized the bipartite fluctuation of the particle number in quench dynamics starting from the alternating state, studying the local equilibrium and the emergence of fluctuating hydrodynamics in the quantum many-body system. 
The experiment focused mainly on the hard-core bosons on a ladder system while also studying the ones on a one-dimensional system, which can be mapped into noninteracting fermions.

Despite the recent strong interest in emergent hydrodynamics in a quantum regime, understanding the experimentally observed bipartite fluctuation of Ref.~\cite{hydro} from the perspective of exact solutions is elusive even in the simplest one-dimensional hard-core bosons \cite{Groha2018,comment1}, although several numerical works~\cite{fujimoto20,Jin20,fujimoto21,fujimoto22,Cecile23,Aditya23,Bhakuni23} exist. Hence, the fundamental understanding of such bipartite fluctuations is highly desirable now, and it is of immense importance to explain the recently observed bipartite fluctuation quantitatively via exact solutions.

In this Letter, we consider one-dimensional fermions starting from the alternating state, theoretically studying time evolution for the variance of the particle numbers in a subsystem as schematically shown in Fig.~\ref{fig0} (a) and compare our theoretical prediction with the recent experimental data of Ref.~\cite{hydro}. First, in the noninteracting fermions, we obtain the exact and simple expression of the variance and derive the asymptotic linear growth of the variance. Second, we compare our analytical result with the experiment of Ref.~\cite{hydro}. For this purpose, we incorporate the incompleteness of the initial alternating state realized in the experiment into our exact solution, deriving the general linear growth law of the variance. This general law agrees well with the variance growth observed in Ref.~\cite{hydro} without any fitting parameters. The impact of the incompleteness is schematically displayed in Fig.~\ref{fig0} (b). Furthermore, utilizing the exact solution before taking the large-size limit, we analytically estimate a time scale for the local equilibration. Finally, we numerically investigate interaction effects on the variance growth using the time-evolving block decimation (TEBD) method~\cite{TEBD1,TEBD2,TEBD3,TEBD4}.

Before ending the introduction, it is worth noting that a bipartite fluctuation has been studied in terms of an integrated current and full counting statistics. 
In the former case, most of previous works consider dynamics starting from a domain-wall initial state, investigating a bipartite fluctuation for a conserved quantity by connecting it to an integrated current via a continuity equation~\cite{Antal2008, Eisler2013, Ljubotina2017, Moriya_2019, Gamayun2020, Jin2021, Gopalakrishnan2023, Wei2022, Krajnik2022_1, Krajnik2022_2, Krajnik2024_1, Krajnik2024_2}. One of the notable results is the logarithmic growth of the variance in noninteracting fermions~\cite{Antal2008}.
In the latter case, a bipartite fluctuation has been employed to study static and dynamic properties of entanglement entropy~\cite{Klich2009,Song2010,Song2012,Stephan2012,Parez2021,Oshima2023,Bertini2023_1} and fluctuations of physical quantities~\cite{Schonhammer2007,Collura2017,Pollmann2017,Najafi2017,Arzamasovs2019,Calabrese2020,Filiberto2021,Smith2021, McCulloch2023, Herce2023,Bertini2023_2}. 

{\it Setup.---}
Let us consider fermions on a one-dimensional lattice labeled by $ \Lambda \coloneqq \{  -L, -L+1, ... , L\} $ with a positive even number $L$. We denote the fermionic creation and annihilation operators at a site $j \in \Lambda $ by $\hat{a}_{j}^{\dagger}$ and $\hat{a}_{j}$. Then, the Hamiltonian is given by
\begin{align} 
\hat{H} \coloneqq -\sum_{j = -L }^{L-1} \left( \hat{a}_{j+1}^{\dagger} \hat{a}_j + \hat{a}_{j}^{\dagger} \hat{a}_{j+1}   \right) + U \sum_{j = -L }^{L-1} \hat{n}_{j+1} \hat{n}_{j}
\label{Hamil}
\end{align}
with the particle-number operator $\hat{n}_{j} \coloneqq \hat{a}_j^{\dagger} \hat{a}_j$ and the interaction parameter $U$. Here, the boundary condition is basically a periodic boundary condition ($\hat{a}_{L} = \hat{a}_{-L}$), 
but we use an open boundary condition ($\hat{a}_{L} = 0$) in several cases. In the latter case, we will explicitly specify the boundary condition. The initial state is the alternating state defined by $\ket{\psi_{\rm alt}} \coloneqq \prod_{j = -L/2 }^{L/2-1} \hat{a}_{2j + 1}^{\dagger} \ket{0}$. We denote a quantum state at time $t$ by $\ket{\psi(t)}$ and assume that the state obeys the Schrödinger equation $ \displaystyle {\rm i} d \ket{\psi(t)}/dt= \hat{H} \ket{\psi(t)}$. Here, the Dirac constant $\hbar$ is set to be unity. 
Note that our model with $U=0$ is equivalent to 1D hard-core bosons. Thus, our theoretical results based on this model are relevant to experimental results for 1D hard-core bosons of Ref.~\cite{hydro}.

In this Letter, we shall study the fluctuation of the particle number in a subsystem. The operator for the bipartite particle-number is defined by $\hat{N}_{M} \coloneqq \sum_{j=0}^{M-1} \hat{a}_j^{\dagger} \hat{a}_j$ with a positive integer $M$. Then, a generating function for the $n$th moment of $\hat{N}_{M}$ is given by $G_M(\lambda,t) \coloneqq \braket{   e^{\lambda \hat{N}_{M} }   }_t$, where we introduce the notation $\braket{  \bullet   }_t \coloneqq {\rm Tr} [\hat{\rho}(t)  \bullet ] $ with the density matrix $\hat{\rho}(t) \coloneqq \ket{\psi(t)} \bra{\psi(t)}$. We can compute the moments by differentiating $G_M(\lambda,t)$ with respect to $\lambda$. 

{\it Exact solution for the variance of the noninteracting fermions.---}
We study the time evolution for the variance of $ \hat{N}_{M} $ in the noninteracting fermions ($U=0$) under the thermodynamic limit ($L \rightarrow \infty$). 
As described in Sec.~I of the Supplemental Material (SM)~\cite{SM}, the generating function $G_M(\lambda,t)$ becomes \cite{Eisler2013,Schonhammer2007,Parez2021}
\begin{align}
G_M (\lambda,t) = {\rm det} \left[ \delta_{j,k} + (e^{\lambda}-1) \braket{ \hat{a}_{j}^{\dagger} \hat{a}_{k} }_t  \right]_{j,k=0}^{M-1}. \label{G1}
\end{align}
Here, the two-point correlator $\braket{ \hat{a}_{j}^{\dagger} \hat{a}_{k} }_t $ in the thermodynamic limit ($L \rightarrow \infty$) is given by~\cite{Flesch2008} 
\begin{align}
\braket{ \hat{a}_{j}^{\dagger} \hat{a}_{k} }_t = \frac{ \delta_{j,k} }{2} - \frac{{\rm i}^{j+k} }{2} J_{k-j}(4t).
\label{C_con}
\end{align}

We here derive the exact and simple expression for the variance of $\hat{N}_M$ under the limit $M \rightarrow \infty$. The first step is to express the variance via the Bessel function $J_n(x)$ of the first kind. Differentiating Eq.~\eqref{G1} with respect to $\lambda$, we obtain the variance $\sigma_M(t)^2 \coloneqq  \braket{ \hat{N}_{M}^2 }_t -  \braket{ \hat{N}_{M} }_t^2 = \partial^2 G_M(\lambda,t)/ \partial \lambda^2|_{\lambda = 0} - (\partial G_M(\lambda,t)/ \partial \lambda|_{\lambda = 0})^2$ as
\begin{align}
\sigma_M(t)^2 = \frac{M}{4} \left( 1 - J_0(4t)^2 - 2 \sum_{k=1}^{M-1} J_k(4t)^2 \right) + \frac{1}{2} \sum_{k=1}^{M-1} k J_{k}(4t)^2. 
\label{variance1}
\end{align}
The detailed derivation of Eq.~\eqref{variance1} is given in Sec.~II of SM~\cite{SM}. 
The next task is to take the limit $M \rightarrow \infty$ in Eq.~\eqref{variance1}. 
As proved in Sec.~III of SM~\cite{SM}, we have
$\lim_{M \rightarrow \infty} M \left( 1 - J_0(4t)^2 - 2 \sum_{k=1}^{M-1} J_k(4t)^2 \right)/4  = 0$ for $t>0$
and
$\lim_{M \rightarrow \infty}  \sum_{k=1}^{M-1} k J_{k}(4t)^2/2 = 4t^2 \left(  J_0(4t)^2 +  J_1(4t)^2 \right) - t  J_0(4t) J_1(4t)$. 
We utilize these formulae and Eq.~\eqref{variance1}, deriving the following exact and simple expression of the variance $\sigma(t)^2 \coloneqq \lim_{M \rightarrow \infty} \sigma_{M}(t)^2 $, 
\begin{align}
\sigma(t)^2 = 4t^2 \left(  J_0(4t)^2 +  J_1(4t)^2 \right) - t  J_0(4t) J_1(4t)
\label{variance7}
\end{align}
for $t>0$. Note that Eq.~\eqref{variance7} is still valid at $t=0$ since we can show $\sigma_{M}(0)^2=0$ and $\sigma(0)^2 = 0$ from Eqs.~\eqref{variance1} and \eqref{variance7}, respectively. Thus, we can relax the constraint $t>0$ for Eq.~\eqref{variance7} to $t \geq 0$.

\begin{figure}[t]
\begin{center}
\includegraphics[keepaspectratio, width=8.7cm]{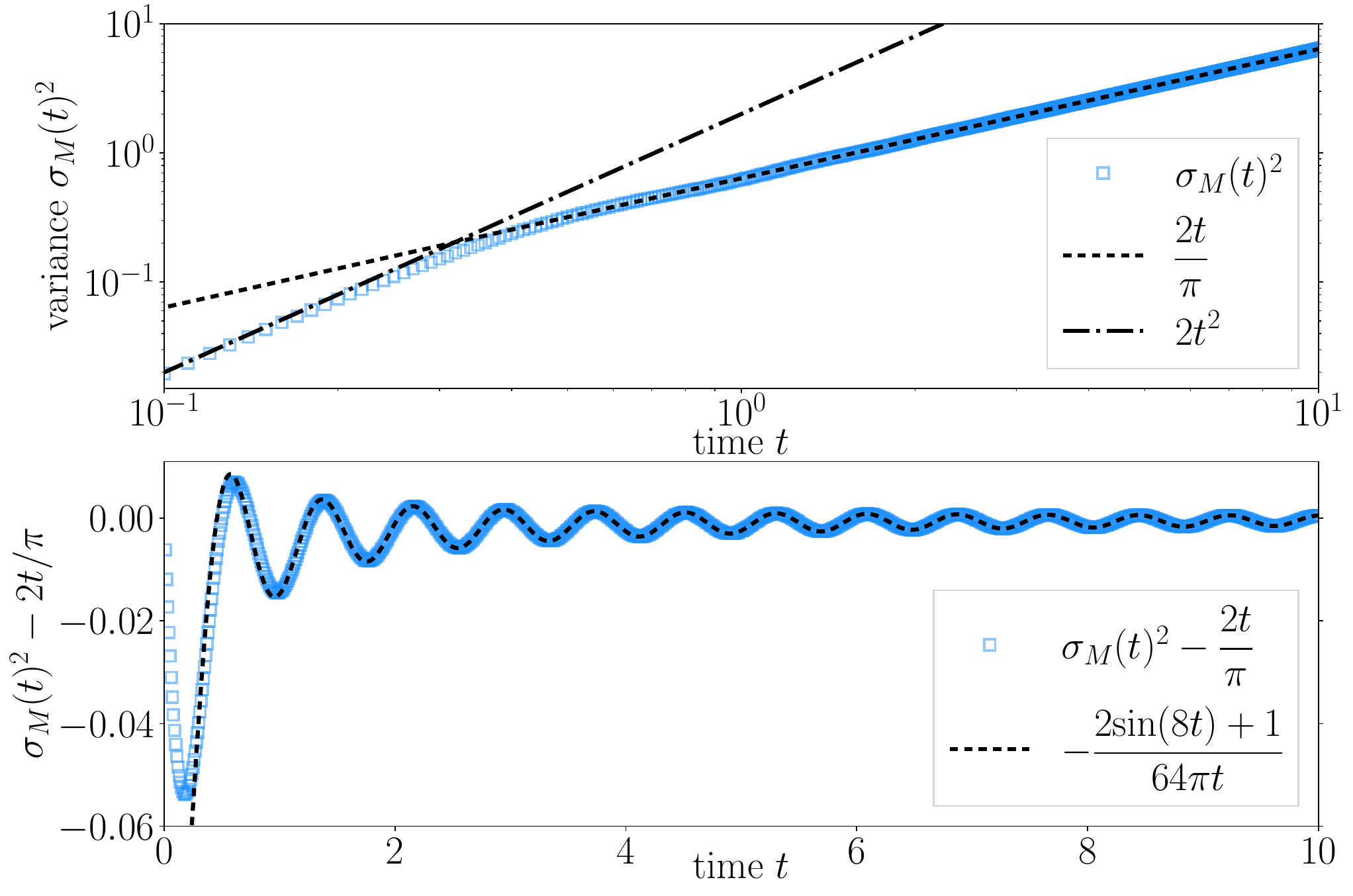}
\caption{
Numerical verification of Eq.~\eqref{asym2}. The time evolution of the variance denoted by the blue square symbols is numerically obtained using Eqs.~\eqref{G1} and \eqref{C_con} with $M=200$. In the upper and lower panels, we compare Eq.~\eqref{asym2} with the numerical result. 
} 
\label{fig1} 
\end{center}
\end{figure}

We apply asymptotic analysis to Eq.~\eqref{variance7}, deriving the asymptotic forms of $\sigma(t)^2$ both for the short-time ($t \ll 1$) and long-time ($t \gg 1$) dynamics. Using the asymptotic forms of the Bessel functions of the first kind~\cite{math1}, we obtain
\begin{align}
\sigma(t)^2 \simeq
 \begin{dcases}
     2 t^2                  & (t \ll 1)     \\
    \frac{2}{\pi} t - \frac{1}{64 \pi t} \left( 2 \sin (8t) + 1 \right)     & (t \gg 1).
  \end{dcases}
  \label{asym2}
\end{align}
The detailed derivation of Eq.~\eqref{asym2} is given in Sec.~IV of SM~\cite{SM}. The essential consequence of Eq.~\eqref{asym2} is that the variance is proportional to time $t$ for $t \gg 1$. This linear growth law is entirely different from the variance growth in noninteracting fermionic dynamics starting from the domain-wall state $\ket{\psi_{\rm DW}} \coloneqq \prod_{j \in \mathbb{Z}_{<0} } \hat{a}_{j}^{\dagger} \ket{0}$, where the variance was shown to obey the logarithmic growth \cite{Antal2008, Moriya_2019}. Note that the linear growth law was numerically confirmed in Ref.~\cite{fujimoto22} and Eq.~\eqref{asym2} gives its analytical explanation.

Finally, we numerically verify Eq.~\eqref{asym2}. Figure~\ref{fig1} shows the time evolution of the variance obtained by Eqs.~\eqref{G1} and \eqref{C_con} numerically. One can see the excellent agreement with Eq.~\eqref{asym2}. Note that the linear growth of the variance appears in $t \gtrsim 0.4$, though it is originally derived under $t \gg 1$. This behavior is favorable for observing the linear growth experimentally.

{\it Comparison with the experimental result of Ref.~\cite{hydro}.---}
We consider whether our theoretical prediction of the variance for the noninteracting fermions can explain the experimental result reported in Ref.~\cite{hydro}. 
For this purpose, we must note that the experiment does not realize the {\it complete} alternating initial state $\ket{\psi_{\rm alt}}$. To make this point clear, let us denote averaged particle numbers at the even and odd sites by $n_{\rm even}$ and $n_{\rm odd}$. The observed imbalance parameter $I \coloneqq (n_{\rm odd} - n_{\rm even} )/(n_{\rm even}  + n_{\rm odd} )$ is about $0.91$~\cite{hydro}, which means the existence of deviation from the {\it complete} alternating state since $\ket{\psi_{\rm alt}}$ has $I=1$. This observation strongly suggests that we need to incorporate the incompleteness of the initial alternating state into the analytical results of Eqs.~\eqref{variance7} and \eqref{asym2}. In what follows, we describe our analytical solution with the {\it incomplete} alternating state, comparing it with the experimental data.

Before calculating the variance, we first specify an initial state for the incomplete alternating state by considering how the alternating state is experimentally prepared. The experiment prepares the initial state by ramping up a strongly tilted double-well potential in one direction (see Sec.~I.~B. of the supplementary information of Ref.~\cite{hydro}). Thus, we expect that the experimental initial state is well described by a product state, assuming that the initial density matrix is approximated by
\begin{align}
\hat{\rho}_{\rm alt} \coloneqq \dfrac{1}{\Xi} \exp \left( -\beta \hat{H}_{\rm alt} + \beta \mu \hat{N}_{\rm tot} \right)
\label{incomp_alt1}
\end{align}
with the inverse temperature $\beta$, the chemical potential $\mu$, and the partition function $\Xi \coloneqq {\rm Tr} \exp ( -\beta \hat{H}_{\rm alt} + \beta \mu \hat{N}_{\rm tot} ) $. The operators $\hat{H}_{\rm alt}$ and $\hat{N}_{\rm tot}$ are defined by $\hat{H}_{\rm alt} \coloneqq \sum_{j=-L/2}^{L/2-1} ( \hat{a}_{2j}^{\dagger} \hat{a}_{2j} - \hat{a}_{2j+1}^{\dagger} \hat{a}_{2j+1}  )$ and $\hat{N}_{\rm tot} \coloneqq \sum_{j=-L}^{L-1} \hat{a}_{j}^{\dagger} \hat{a}_{j}$. The averaged particle numbers at the even and odd sites become $n_{\rm even} = 1/[ e^{ \beta(1  -  \mu)} + 1 ]$ and $n_{\rm odd} = 1/[ e^{ \beta(-1 - \mu)} + 1 ]$. The parameters $\beta$ and $\mu$ are determined by the filling factor $\nu \coloneqq (n_{\rm even}+n_{\rm odd})/2$ and the imbalance parameter $I$, both of which are observable in the experiment.

We analytically derive the exact and asymptotic forms of the variance for the dynamics starting from the incomplete alternating state of Eq.~\eqref{incomp_alt1}.
As detailed in Sec.~V of SM~\cite{SM}, the two-point correlator becomes
\begin{align}
\braket{ \hat{a}_{j}^{\dagger} \hat{a}_{k} }_t = \frac{1}{2} \left( n_{\rm even} + n_{\rm odd}  \right) \delta_{j,k} + \frac{1}{2}  \left( n_{\rm even} - n_{\rm odd}  \right) {\rm i}^{j+k} J_{k-j}(4t)
\label{incomp_alt4}
\end{align}
in the thermodynamic limit ($L \rightarrow \infty$). Here the density matrix is given by $\hat{\rho}(t) =  e^{ -{\rm i} \hat{H}t } \hat{\rho}_{\rm alt} e^{ {\rm i} \hat{H}t } $ with Eq.~\eqref{Hamil} and $U=0$.
Following the derivation given in Sec. VI of SM~\cite{SM}, we obtain the variance as 
\begin{widetext}
\begin{align}
\sigma_{M}(t)^2  = \delta \sigma_M(t)^2 + (n_{\rm even} - n_{\rm odd})^2 \left[ \dfrac{M}{4} \left( 1 - J_0(4t)^2 - 2 \sum_{m=1}^{M-1}J_m(4t)^2 \right) + \dfrac{1}{2} \sum_{m=1}^{M-1} m J_m(4t)^2 \right].
\label{sigma_M}
\end{align}
\end{widetext}
with the function $ \delta \sigma_{M}(t)^2 \coloneqq M [ n_{\rm even}(1-n_{\rm even}) + n_{\rm odd}(1-n_{\rm odd})]/2 + J_{0}(4t)[ n_{\rm even}(1-n_{\rm even}) - n_{\rm odd}(1-n_{\rm odd})] [\sum_{m=0}^{M-1}(-1)^m]/2$. Under the setup, we can derive
\begin{align}
&\sigma_{\rm sub}(t)^2 \coloneqq \lim_{M \rightarrow \infty} \left( \sigma_{M}(t)^2 - \delta \sigma_{M}(t)^2  \right) \nonumber\\
&= \left( n_{\rm even} - n_{\rm odd}  \right)^2 \Bigl[ 4t^2 \left(  J_0(4t)^2 +  J_1(4t)^2 \right) -t  J_0(4t) J_1(4t) \Bigl]
\label{incomp_alt5}
\end{align}
Here, the time-dependent term of $\delta \sigma_{M}(t)^2$ is much smaller than the time-independent term when $M$ is large. Hence, we have $\delta \sigma_{M}(t)^2  \simeq \sigma_{M}(0)^2 $ for $M \gg 1$. 
Thus $\sigma_{\rm sub}(t)^2$ can be interpreted as the variance from which one subtracts its initial value. Applying the asymptotic analysis used in Eq.~\eqref{asym2} to Eq.~\eqref{incomp_alt5}, we derive
\begin{align}
\sigma_{\rm sub}(t)^2 \simeq
 \begin{dcases}
     2 \left( n_{\rm even} - n_{\rm odd}  \right)^2 t^2                  & (t \ll 1)     \\
    \dfrac{2}{\pi} \left( n_{\rm even} - n_{\rm odd}  \right)^2 t     & (t \gg 1).
  \end{dcases}
  \label{incomp_alt7}
\end{align}
This result elucidates that the incompleteness of the alternating state substantially affects the coefficient of the variance, but the exponents of time are robust.  

\begin{figure}[t]
\begin{center}
\includegraphics[keepaspectratio, width=8.7cm]{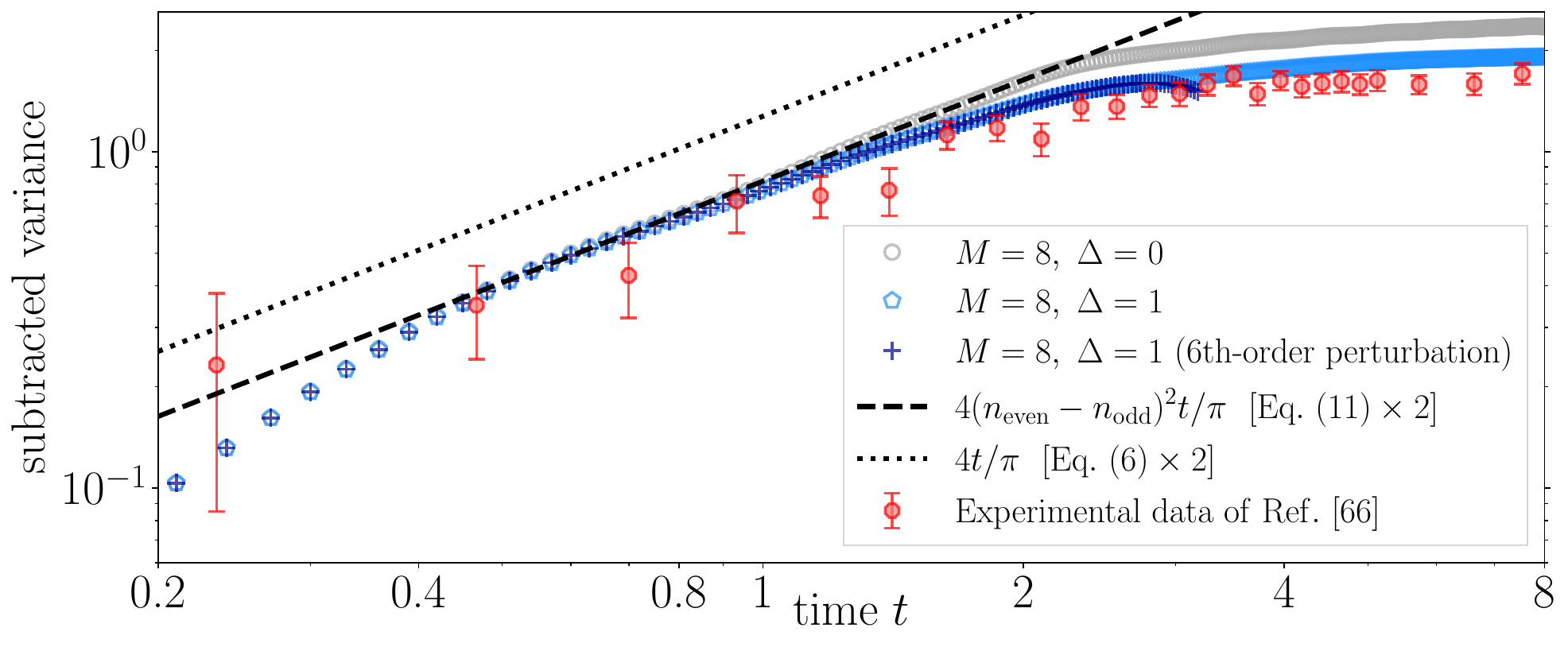}
\caption{
Comparison of our theoretical prediction with the experimental data. The parameters are $2L=40$, $\nu = 0.44$, $I = 0.91$, and  $M=8$, which are taken from Ref.~\cite{hydro}. The time evolution of the variance, denoted by circle and pentagon markers, is numerically obtained using Eq.~\eqref{Hamil_exp} with $\Delta = 0$ and $1$. For the comparison with the experiment, we plot the subtracted variance $2\sigma'_M(t)^2 - 2\sigma'_M(0)^2$ for the theoretical results and $\sigma_{\rm exp}(t)^2 - \sigma_{\rm exp}(t_0)^2$ for the experimental result because the experiment studies two-ladder lattice systems and thus the observed variance is twice as large as ours. Here, the experimental data $\sigma_{\rm exp}(t)^2$ is taken from Fig.~S8 of Ref.~\cite{hydro}, and $t_0$ is $0.0006031857894892$. The dotted and dashed lines correspond to our analytical expressions of Eqs.~\eqref{asym2} and \eqref{incomp_alt7} for $t \gg1$, respectively. The plus marker denotes the numerical data of the 6th-order perturbative calculation. 
}  
\label{fig3} 
\end{center}
\end{figure}

In addition to Eq.~\eqref{incomp_alt7}, we shall compare the numerical calculations in a finite system with the experiment. Our numerical calculation is implemented for $2L=40$, $\nu = 0.44$, and $I = 0.91$, which are taken from Ref.~\cite{hydro} (see Sec.~II.~C. of the supplementary information of Ref.~\cite{hydro}). We adopt the Hamiltonian defined by 
\begin{align} 
\hat{H}' \coloneqq -\sum_{j = -L }^{L-1} \left( \hat{a}_{j+1}^{\dagger} \hat{a}_j + \hat{a}_{j}^{\dagger} \hat{a}_{j+1}   \right) +  \sum_{j = -L }^{L-1} V_j \hat{n}_{j}
\label{Hamil_exp}
\end{align}
with the open boundary condition ($\hat{a}_L=0$). Following the numerical simulation in Ref.~\cite{hydro}, we here add a random potential $V_j$ sampled from a uniform distribution with the range $[ -\Delta, \Delta]$. Here, $\Delta \geq 0$ denotes the strength of the randomness. Under this setup, we investigate the variance $\sigma'_M(t)^2 \coloneqq \overline{ \braket{\hat{N}_M'^2 } } -  \left( ~\overline{ \braket{\hat{N}_M' } }~  \right)^2$ with $\hat{N}_{M}' \coloneqq \sum_{m=-M/2}^{M/2-1} \hat{a}^{\dagger}_m \hat{a}_m$. Here, the overline $\overline{\bullet}$ denotes the ensemble average over the random potentials, and we use 10000 samples for this. 

\begin{figure}[t]
\begin{center}
\includegraphics[keepaspectratio, width=8.7cm]{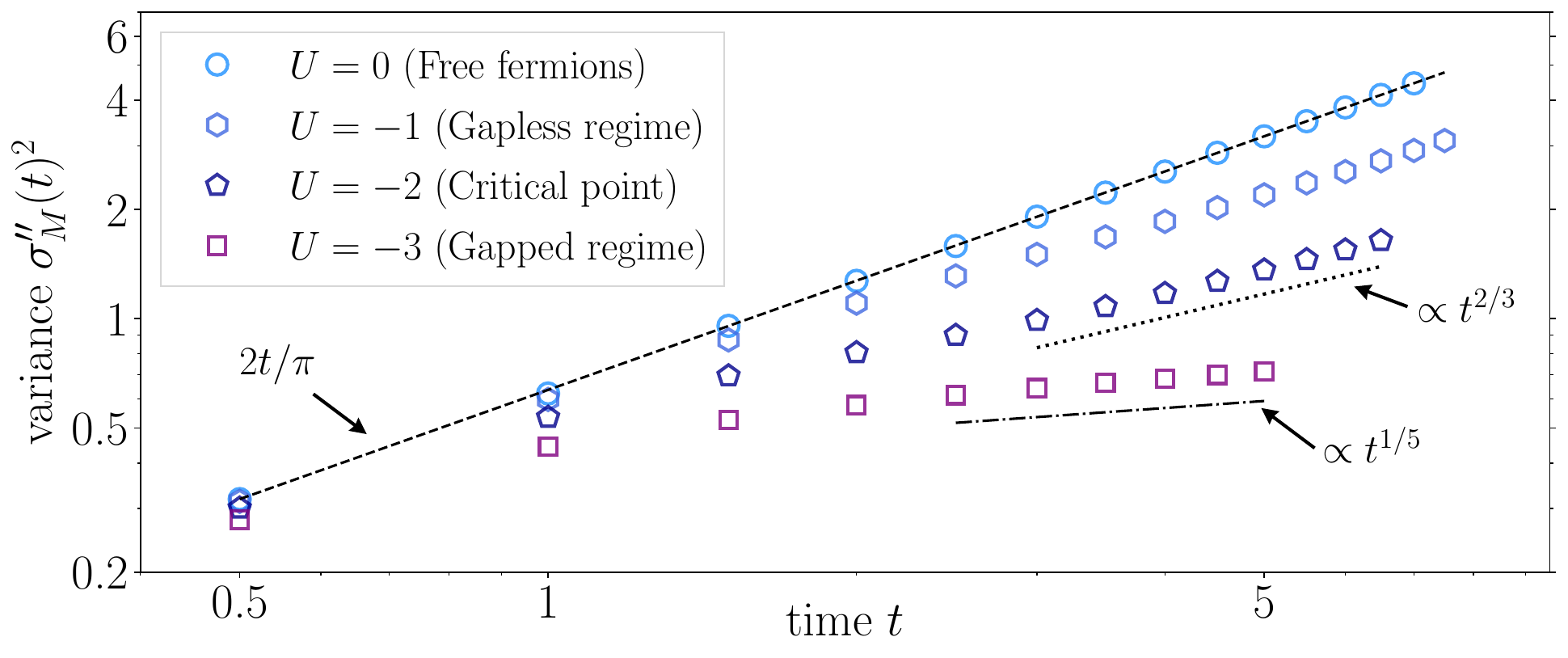}
\caption{
Numerical results for the variance $\sigma_{M}''(t)^2$ with $M=40$. The number of the lattice is $2L=120$. We numerically solve the Schrödinger equation using the TEBD method~\cite{TEBD1,TEBD2,TEBD3,TEBD4}, calculating the variances for $U = 0, -1, -2$, and $-3$. The dashed line denotes the leading term of Eq.~\eqref{asym2} for $t \gg 1$, and the dotted and dot-dashed lines are curves for $t^{2/3}$ and $t^{1/5}$, respectively. We insert these curves with the exponents $2/3$ and $1/5$ for reference. 
}  
\label{fig2} 
\end{center}
\end{figure}

Figure~\ref{fig3} compares our theoretical results with the experimental data of Ref.~\cite{hydro}. Our numerical result with $\Delta = 0$ can capture the variance growth in the early stage of the dynamics, but the deviation from the experimental data grows in time. On the other hand, when we turn on the random potential with $\Delta = 1$, the numerical result well reproduces the experimental data. The saturation of the variance for $t \gtrsim 2$ is caused by the finite size effect and the disorder potential. These behaviors were reported in the numerical simulations of Ref.~\cite{hydro}. We also implement a numerical perturbative calculation with $\Delta$, finding that the 6th-order perturbative result well describes the time evolution before the saturation (see Sec.~VII of SM for the detailed numerical method). Comparing these results, one can see that the disorder effect is irrelevant in $t \lesssim 1$. 

The dotted and dashed lines in Fig.~\ref{fig3} show the analytical results for the linear growth of Eqs.~\eqref{asym2} and \eqref{incomp_alt7}. We find that the latter including the incompleteness of the initial alternating state exhibits the good agreement with the experimental linear growth, while the former not including the incompleteness systematically deviates from the experimental data. We here emphasize that our analytical result~\eqref{incomp_alt7} agrees well with the experimental data without any fitting parameters under the assumption of \eqref{incomp_alt1}. Note that one can see that the disorder effect emerges before the finite $M$ effect by comparing the numerical data for $\Delta=0$ and 1 with Eq.~\eqref{incomp_alt7}, though the localization length is larger than $M=8$~\cite{hydro}. 

We comment on the fact that our incomplete initial density matrix of Eq.~\eqref{incomp_alt1} cannot describe superpositions of quantum states between different sites. The experiment of Ref. ~\cite{hydro} is expected to suppress such entangled states since deep optical lattices are used for the initial state preparation, but we numerically study the effects of the superpositions on the variance growth by extending Eq.~\eqref{incomp_alt1} to a more general state. As illustrated in Sec. VIII of SM~\cite{SM}, we numerically confirm that the results of Eq.~\eqref{incomp_alt1} show better agreement with the experiment than ones of the general state. This finding strengthens the justification for using Eq.~\eqref{incomp_alt1}.   

Finally, we discuss a time scale for local equilibration on the basis of our exact solution of Eq.~\eqref{sigma_M} for a finite $M \gg 1$, where the linear growth of the subtracted variance, namely $2 \left( n_{\rm even} - n_{\rm odd}  \right)^2 t / \pi$, becomes dominant (see Eq.~\eqref{incomp_alt7}). When taking the long-time limit of Eq.~\eqref{sigma_M}, we obtain the stationary value of the subtracted variance $\lim_{t \rightarrow \infty} (\sigma_{M}(t)^2  - \delta \sigma_M(t)^2) =  (n_{\rm even} - n_{\rm odd})^2 M/4$. Then, we can estimate the local equilibration time $T_{*}$ by equating the expressions for these two subtracted variances, finding $T_* \sim \pi M /8 $. Interestingly, our exact solution leads to the fact that $T_*$ is independent of the initial value of $n_{\rm even} - n_{\rm odd}$.  

{\it Numerical study for the variance growth of the interacting fermions.---}
We numerically study the interaction effect on the time evolution of the variance. Our numerical method is the TEBD method~\cite{TEBD1,TEBD2,TEBD3,TEBD4} with Eq.~\eqref{Hamil} and the open boundary condition ($\hat{a}_L=0$). We set $L$ to be $60$, and compute the variance for $U = 0, -1, -2$, and $-3$. In the language of the XXZ chain, $U = -2$ corresponds to the critical point at which the model is identical to the XXX chain~\cite{franchini2017}. To weaken the boundary effect, we use the variance $\sigma''_M(t)^2 \coloneqq \langle ( \hat{N}_{M}')^2 \rangle_t - \langle \hat{N}_{M}' \rangle_t^2$ with $M=40$. 
The dependence on the bond dimensions and the time resolutions of our numerical calculations is described in Sec. X of SM~\cite{SM}.

Figure~\ref{fig2} displays the time evolution of the variance $\sigma''_M(t)^2$. In the noninteracting fermions ($U=0$), our numerical result for the finite system well reproduces the leading term of Eq.~\eqref{asym2} for the infinite system. This demonstrates that the boundary effect is negligible. In the interacting cases ($U \neq 0$), the time evolution exhibits different growth behaviors, and we cannot find the clear ballistic property, particularly for the $U=-2$ and $-3$ cases. 

Note that Cecile {\it et al.} recently reported similar numerical results for the variance in the XXZ chain \cite{Cecile23}. 
For example, they confirmed the signature of the anomalous power law growth ($\sigma''_{M}(t)^2 \propto t^{2/3}$) in the dynamics starting from the N$\rm \acute{e}$el state identical to the alternating state. The similar signature was also discussed in Ref.~\cite{fujimoto20}. We here display Fig.~\ref{fig2} to show that our analytical result for the noninteracting fermions does not work in the interacting cases. 

Finally, we study the effect of the random potential on the variance growth. We numerically find that the disorder effect emerges in the early stage of the dynamics and the power-law-like growth in Fig.~\ref{fig2} becomes unclear, as described in Sec. IX of SM~\cite{SM}. This numerical finding suggests that one needs to decrease the strength of the random potential to observe the power-law-like growth in the interacting fermions, namely the XXZ model. 

{\it Future prospects.---}
Our theoretical results, based upon the exact solutions of the noninteracting fermions and the numerical findings of the interacting fermions will stimulate further research on emergent hydrodynamics and local equilibration of quantum many-body systems. As a prospect, it is fascinating to analytically comprehend the interaction effect via generalized hydrodynamics \cite{Olalla2016, Bertini2016, Bulchandani2017, Bulchandani18, Doyon2017, Doyon2017_2, Doyon2018, Collura2018, Jacopo2018, Schemmer2019, Sarang2019, Doyon2020_rev, Alba2021, Malvania2021, Bouchoule2022, Essler2023} and ballistic macroscopic fluctuation theory \cite{BMA1,BMA2}. In another direction, it is worth studying the variance growth in open quantum systems. Depending on the kinds of interactions with environments, an open quantum system approaches a nonequilibrium stationary state being different from the equilibrium state addressed in this work. Thus, uncovering features of the variance growth in such a case is fundamentally intriguing. 

\begin{acknowledgments}
KF is grateful to Ryusuke Hamazaki and Yuki Kawaguchi for fruitful discussions and comments on the manuscript, Masaya Kunimi for helpful discussions on TEBD with a conserved quantity, and Hiroki Moriya for helpful discussions. KF and ST are grateful to Monika Aidelsburger, Immanuel Bloch, Alexander Impertro, Simon Karch, Christian Schweizer, and Julian F. Wienand for sharing the experimental data of Ref.~\cite{hydro} used in Fig.~\ref{fig3} of the main text and Figs.~\ref{Sfig3} and \ref{Sfig4} of SM~\cite{SM}. 
The work of KF has been supported by JSPS KAKENHI Grant No. JP23K13029. 
The work of TS has been supported by JSPS KAKENHI Grants No. JP21H04432, No. JP22H01143.
\\

{\it Note added.---}
After uploading our manuscript on arXiv, O. Gamayun have reported that the bipartite fluctuations growing from the alternating state is related to that from the domain-wall state \cite{Gamayun}. 
\end{acknowledgments}
\bibliography{reference}

\widetext
\clearpage

\setcounter{equation}{0}
\setcounter{figure}{0}
\setcounter{section}{0}
\setcounter{table}{0}
\renewcommand{\theequation}{S-\arabic{equation}}
\renewcommand{\thefigure}{S-\arabic{figure}}
\renewcommand{\thetable}{S-\arabic{table}}

\section*{Supplemental Material for ``Exact solution of bipartite fluctuations in one-dimensional fermions''}

\centerline{Kazuya Fujimoto and Tomohiro Sasamoto}
\vspace{3mm}

\centerline{Department of Physics, Institute of Science Tokyo, 2-12-1 Ookayama, Meguro-ku, Tokyo 152-8551, Japan}
\vspace{5mm}

\par\vskip3mm \hrule\vskip.5mm\hrule \vskip.30cm
This Supplemental Material describes the following:
\begin{itemize}
\item[  ]{ (I) Derivation of Eq.~\eqref{G1}, } 
\item[  ]{ (II) Derivation of Eq.~\eqref{variance1}, } 
\item[  ]{ (III) Derivation of the two limiting formulae, } 
\item[  ]{ (IV) Derivation of Eq.~\eqref{asym2}, } 
\item[  ]{ (V) Derivation of Eq.~\eqref{incomp_alt4}, } 
\item[  ]{ (VI) Derivation of Eq.~\eqref{incomp_alt5}, } 
\item[  ]{ (VII) Numerical perturbative calculation for the disorder potential, } 
\item[  ]{ (VIII) Effect of the superposition of the initial state in the noninteracting fermions, } 
\item[  ]{ (IX) Effect of the random potential on the interacting fermions, } 
\item[  ]{ (X) Dependence on the bond dimension and the time resolution in the numerical calculations. }

\end{itemize}
\par\vskip1mm \hrule\vskip.5mm\hrule

\vspace{5mm}

\section{Derivation of Eq.~(\ref{G1})}
We derive the determinantal formula for the generating function, namely Eq.~(\ref{G1}) of the main text.
The essential ingredient of the derivation is the fact that the Wick theorem can be applicable since the quantum state in our setup is the Gaussian state.  
The detailed calculation is given by
\begin{align}
G_M (\lambda,t) &= \left\langle \prod_{j=0}^{M-1} e^{ \lambda \hat{a}_{j}^{\dagger}\hat{a}_{j} } \right\rangle_t  \nonumber\\
&= \left\langle \prod_{j=0}^{M-1} \biggl[  1 + (e^{\lambda}-1) \hat{a}_{j}^{\dagger}\hat{a}_{j} \biggl] \right\rangle_t \nonumber\\
&= 1 + \sum_{n=1}^{M} (e^{\lambda}-1)^n \sum_{ \substack{j_1 < j_2 < ... < j_n \\  j_{k} \in \{0,1,2,...,M-1\} } }\left\langle \prod_{k=1}^{n} \hat{a}_{j_{k}}^{\dagger}\hat{a}_{j_{k}}  \right\rangle_t \nonumber\\
&= 1 + \sum_{n=1}^{M} (e^{\lambda}-1)^n \sum_{ \substack{j_1 < j_2 < ... < j_n \\  j_{k} \in \{0,1,2,...,M-1\} } } {\rm det} \left[ \braket{ \hat{a}_{j_k}^{\dagger} \hat{a}_{j_l} }_t  \right]_{k,l=1}^{n} \nonumber\\
&= {\rm det} \left[ \delta_{j,k} + (e^{\lambda}-1) \braket{ \hat{a}_{j}^{\dagger} \hat{a}_{k} }_t  \right]_{j,k=0}^{M-1}.
\end{align}
In the fourth line, we use the Wick theorem. This kind of determinantal form was derived in several previous works~\cite{Schonhammer2007,Eisler2013,Parez2021}.

\section{Derivation of Eq.~(\ref{variance1})}\label{sec:1}
\begin{figure}[t]
\begin{center}
\includegraphics[keepaspectratio, width=10.0cm]{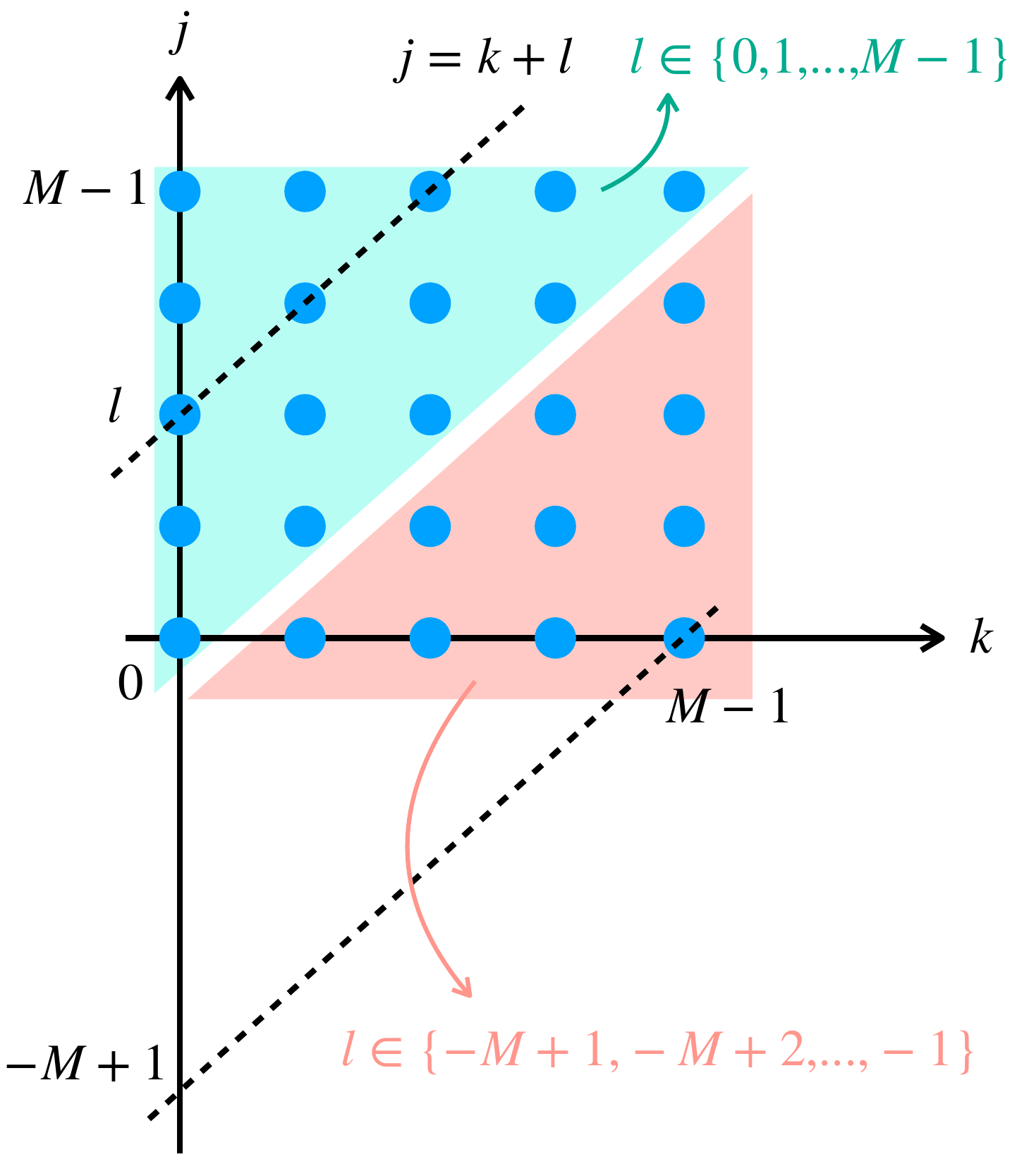}
\caption{
Schematic illustration in deriving Eq.~\eqref{App:A2_2}. The filled circles denotes the points for the double summation $\sum_{j=0}^{M-1} \sum_{k=0}^{M-1} \bullet$. In the blue region specified by the set $\{ (k,j) \in \{0,1,..., M-1 \}^2 \mid j \geq k\}$, the $j$-intercept $l = j-k$ can take the values from $\{ 0,1,...,M-1 \}$. On the other hand, the $j$-intercept takes the values from $\{-M+1, -M+2 ,..., -1 \}$ in the red region specified by the set $\{ (k,j) \in \{0,1,..., M-1 \}^2 \mid  j < k\}$. Using the $j$-intercept $l$, we can transform the double summation of the left-hand side of Eq.~\eqref{App:A2_2} into the right-hand side of Eq.~\eqref{App:A2_2}.
}  
\label{Sfig1} 
\end{center}
\end{figure}

We express the variance $\sigma_M(t)^2$ by the Bessel function $J_n(x)$ of the first kind. Using the property of the generating function, we can obtain
\begin{align}
\sigma_M(t)^2 &= \left. \dfrac{\partial ^2}{\partial \lambda^2} G_M(\lambda,t) \right|_{\lambda = 0} - \left( \left. \dfrac{\partial}{\partial \lambda} G_M(\lambda,t) \right|_{\lambda = 0} \right)^2 \label{variance0}  \\
&= \sum_{k=0}^{M-1} \braket{ \hat{a}_k^{\dagger} \hat{a}_k }_t  - \sum_{j=0}^{M-1} \sum_{k=0}^{M-1} | \braket{ \hat{a}_j^{\dagger} \hat{a}_k }_t |^2.
\label{S_variance1}
\end{align}
The first term on the right-hand side of Eq.~\eqref{S_variance1} becomes
\begin{align}
\sum_{k=0}^{M-1} \braket{ \hat{a}_k^{\dagger} \hat{a}_k }_t  = \frac{M}{2}  - \frac{ J_{0}(4t) }{2} \sum_{k=0}^{M-1} (-1)^{k}, 
\label{S_variance2}
\end{align}
where we use the following expression of the two-point correlator, 
\begin{align}
\braket{ \hat{a}_{j}^{\dagger} \hat{a}_{k} }_t = \frac{ \delta_{j,k} }{2} - \frac{{\rm i}^{j+k} }{2} J_{k-j}(4t).
\label{S_C_con}
\end{align}
Similarly, the second term on the right-hand side of Eq.~\eqref{S_variance1} becomes
\begin{align}
\sum_{j=1}^{M-1} \sum_{k=1}^{M-1} | \braket{ \hat{a}_j^{\dagger} \hat{a}_k }_t |^2  &=  \frac{M}{4} - \frac{ J_0(4t) }{2} \sum_{k=0}^{M-1} (-1)^k  + \frac{1}{4} \sum_{j=0}^{M-1} \sum_{k=0}^{M-1} J_{j-k}(4t)^2.
\label{App:A1}
\end{align}
The double summation on the right-hand side of Eq.~\eqref{App:A1} becomes
\begin{align}
\sum_{j=0}^{M-1} \sum_{k=0}^{M-1} J_{j-k}(4t)^2 & = \sum_{l=0}^{M-1} \sum_{m=0}^{M-l-1} J_{l}(4t)^2 + \sum_{l=-M+1}^{-1} \sum_{m=-l}^{M-1} J_{l}(4t)^2 \label{App:A2_2} \\
& = M   J_{0}(4t)^2  + 2\sum_{l=1}^{M-1} \left( M - l  \right) J_{l}(4t)^2 \label{App:A2_3} \\
& = M  \left( J_{0}(4t)^2 + 2 \sum_{l=1}^{M-1} J_{l}(4t)^2   \right) -  2 \sum_{l=1}^{M-1} l J_{l}(4t)^2. \label{App:A2_4}
\end{align}
In Eq.~\eqref{App:A2_2}, we introduce $l = j - k$ and divide the single double summation into the two double summations. 
Figure~\ref{Sfig1} schematically describes this procedure in detail. 
We also use the formula $J_{-l}(4t)^2 = J_{l}(4t)^2$ in Eq.~\eqref{App:A2_3}.
Finally, putting Eq.~\eqref{App:A2_4} into Eq.~\eqref{App:A1}, we get 
\begin{align}
\sum_{j=1}^{M-1} \sum_{k=1}^{M-1} | \braket{ \hat{a}_j^{\dagger} \hat{a}_k }_t |^2  &=  \frac{M}{4} - \frac{ J_0(4t) }{2} \sum_{k=0}^{M-1} (-1)^k  + \frac{M}{4}  \left( J_{0}(4t)^2 + 2 \sum_{l=1}^{M-1} J_{l}(4t)^2   \right) -  \dfrac{1}{2} \sum_{l=1}^{M-1} l J_{l}(4t)^2.
\label{App:A3}
\end{align}
Substituting Eqs.~\eqref{S_variance2} and \eqref{App:A3} into \eqref{S_variance1}, we derive Eq.~(\ref{variance1}) of the main text, namely
\begin{align}
\sigma_M(t)^2 = \frac{M}{4} \left( 1 - J_0(4t)^2 - 2 \sum_{k=1}^{M-1} J_k(4t)^2 \right) + \frac{1}{2} \sum_{k=1}^{M-1} k J_{k}(4t)^2. 
\label{App:A4}
\end{align}

\section{Derivation of the two limiting formulae }
In the derivation of Eq.~\eqref{variance7}, we use the following formulae, 
\begin{align}
& \lim_{M \rightarrow \infty} \frac{M}{4} \left( 1 - J_0(4t)^2 - 2 \sum_{k=1}^{M-1} J_k(4t)^2 \right)  = 0~~~(t>0),  \label{S_lim1}\\
& \lim_{M \rightarrow \infty} \frac{1}{2} \sum_{k=1}^{M-1} k J_{k}(4t)^2 = 4t^2 \left(  J_0(4t)^2 +  J_1(4t)^2 \right) - t  J_0(4t) J_1(4t) \label{S_lim2}.
\end{align}
In this section, we shall prove them. 

\subsection{ Proof of Eq.~(\ref{S_lim1})  }
For the proof of Eq.~\eqref{S_lim1}, we first derive the following inequality, 
\begin{align}
\frac{M}{4} \left|  \left( 1 - J_0(4t)^2 - 2 \sum_{k=1}^{M-1} J_k(4t)^2 \right)  \right|  < \dfrac{1}{2} M^{(\gamma-1)M+1} (M+1)^{(\gamma-1)(M+1)+1} (2t)^{2M}
\label{App:B1}
\end{align}
for $t>0$. Here $\gamma \simeq 0.577$ is the Euler-Mascheroni constant. 
The proof of Eq.~\eqref{App:B1} is described in the following. We employ the recursion relation $2  d J_n(x)/dx = J_{n-1}(x) - J_{n+1}(x)$ for the Bessel function of the first kind, deriving
\begin{align}
\frac{d}{dx} \sum_{n=1}^{M-1} J_n(x)^2 &= \sum_{n=1}^{M-1} \left( J_n(x) J_{n-1}(x) - J_n(x) J_{n+1}(x) \right) \\
& = J_1(x) J_0(x) - J_{M-1}(x) J_{M}(x) \\
& = - \frac{1}{2} \frac{d}{dx} J_0(x)^2 - J_{M-1}(x) J_{M}(x),
\label{App:B2}
\end{align}
where we use the formula $J_{1}(x) = - d J_0(x)/dx$ in the last equality. Integrating Eq.~\eqref{App:B2} from $0$ to $x$ leads to
\begin{align}
J_{0}(x)^2 + 2 \sum_{k=1}^{M-1} J_{k}(x)^2 - 1 = - 2\int_{0}^{x} J_{M-1}(y) J_{M}(y) dy, 
\label{App:B3}
\end{align}
where we use $J_{0}(0) = 1$ and $J_{l}(0) = 0~(l >0)$. We next consider the absolute value of Eq.~\eqref{App:B3} and then obtain 
\begin{align}
&\left| \left( 1 - J_0(x)^2 - 2 \sum_{k=1}^{M-1} J_k(x)^2 \right)  \right| \\
&\leq  2\int_{0}^{x} \left| J_{M-1}(y) \right|  \left| J_{M}(y)\right|  dy \\
& < 2 M^{(\gamma-1)M + 1} (M+1)^{(\gamma-1)(M+1) + 1}  \int_0^x \left( \frac{y}{2} \right)^{2M-1} \exp\left( - \frac{y^2}{4M} - \frac{y^2}{4M+4} \right) dy  \label{App:B3_3} \\
& < 2 M^{(\gamma-1)M + 1} (M+1)^{(\gamma-1)(M+1) + 1}  \int_0^x \left( \frac{y}{2} \right)^{2M-1}  dy \\
& = 2M^{(\gamma-1)M} (M+1)^{(\gamma-1)(M+1) + 1} \left( \frac{x}{2} \right)^{2M}. \label{App:B3_5}
\end{align}
In Eq.~\eqref{App:B3_3}, we use the following inequality, 
\begin{align}
J_{M}(x) < \left( \frac{x}{2} \right)^M (M+1)^{(\gamma-1)(M+1)+1} \exp \left( - \frac{x^2}{4M + 4} \right)
\end{align}
for $x>0$. This inequality can be derived using the two inequalities given by 
\begin{align}
J_{\nu}(x) < \left( \frac{x}{2} \right)^{\nu} \frac{1}{\Gamma(\nu+1)} \exp \left( - \frac{x^2}{4\nu + 4} \right)
\label{App:B4}
\end{align}
for $\nu \geq 0 $ and $x > 0$, and 
\begin{align}
x^{(1-\gamma)x - 1} <  \Gamma(x) < x^{x-1}
\label{App:B5}
\end{align}
for $x \in (1,\infty)$. Here, $\Gamma(x)$ denotes the Gamma function. The inequalities of Eqs.~\eqref{App:B4} and \eqref{App:B5} were given in Refs.~\cite{Watson1922, Ifantis1990} and \cite{Andersons1997}, respectively. 
Thus, employing the inequality of Eq.~\eqref{App:B3_5} with $x = 4t$, we can readily derive the inequality of Eq.~\eqref{App:B1}.

Finally, we take the limit $M \rightarrow \infty$ in Eq.~\eqref{App:B1}. For this purpose, we consider the logarithmic of the right-hand side of Eq.~\eqref{App:B1}, which reads
\begin{align}
 -\log2 + \left[ (\gamma-1)M + 1\right] \log M  +  \left[ (\gamma-1)(M+1) + 1\right] \log (M+1)  + 2M \log(2t).
\label{App:B6}
\end{align}
Thus, when $M$ is much larger than unity, the above for a fixed nonzero time $t$ can be approximated to be $2(\gamma-1)M \log M$. This becomes $-\infty$ for $M \rightarrow \infty$ because the Euler-Mascheroni constant $\gamma  \simeq 0.577$ is smaller than unity. Hence, the left-hand side of Eq.~\eqref{App:B1} for $M \rightarrow \infty$ can be bounded by zero. This proves Eq.~\eqref{S_lim1}.

\subsection{ Proof of Eq.~(\ref{S_lim2})  }
We shall prove Eq.~\eqref{S_lim2} by noting the recursion relation $2k J_k(x)/x = J_{k-1}(x) + J_{k+1}(x)$ for the Bessel function of the first kind. This relation leads to 
\begin{align}
& \frac{2k}{x}J_{k}(x) J_{k}(y) = J_{k-1}(x) J_{k}(y) + J_{k+1}(x) J_{k}(y), \\
& \frac{2k}{y}J_{k}(x) J_{k}(y) = J_{k-1}(y) J_{k}(x) + J_{k+1}(y) J_{k}(x).
\label{App:C1}
\end{align}
Subtracting them and taking the summation for the index $k$ from zero to a positive integer $N$, we obtain
\begin{align}
 \left( \frac{2}{x} - \frac{2}{y} \right) \sum_{k=0}^{N} k J_k(x) J_k(y) & = \sum_{k=1}^{N} ( J_{k+1}(x) J_{k}(y) - J_{k+1}(y) J_{k}(x)  + J_{k-1}(x) J_{k}(y) - J_{k-1}(y) J_{k}(x) ) \\
\nonumber \\
& = J_{N+1}(x) J_{N}(y) - J_{N+1}(y) J_{N}(x)  + J_{0}(x) J_{1}(y) - J_{0}(y) J_{1}(x), 
\label{App:C2}
\end{align}
where we use the property of the telescoping series to derive the last expression. Taking the limit $N \rightarrow \infty$, we get
\begin{align}
\sum_{k=0}^{\infty} k J_k(x) J_k(y) = \frac{xy}{2} \frac{ J_{0}(x) J_{1}(y) - J_{0}(y) J_{1}(x) }{y-x}. 
\label{App:C3}
\end{align}
Finally, we take the limit $y \rightarrow x$ in Eq.~\eqref{App:C3} via the L'Hôpital's rule, obtaining 
\begin{align}
\sum_{k=0}^{\infty} k J_k(x)^2= \frac{x^2}{2} J_0(x)^2 + \frac{x^2}{2} J_1(x)^2 - \frac{x}{2} J_0(x) J_1(x).
\label{App:C4}
\end{align}
Here, we use the following formulae, 
\begin{align}
& \frac{d}{dx} J_{k+1}(x) = J_k(x) - \frac{k+1}{x} J_{k+1}(x), \\
& \frac{d}{dx} J_{k-1}(x) = -J_k(x) + \frac{k-1}{x} J_{k-1}(x).
\label{App:C5}
\end{align}
Finally, we put $x=4t$ into Eq.~\eqref{App:C4}, completing the proof of Eq.~\eqref{S_lim2}.

\section{ Derivation of Eq.~(\ref{asym2})}
We derive the asymptotic expression of the variance given by Eq.~\eqref{asym2} of the main text for $t \ll 1$ and $t \gg 1$.
For the short-time dynamics ($t \ll 1$), the Bessel functions $J_0(4t)$ and $J_1(4t)$ of the first kind is approximated by
\begin{align}
&J_{0}(4t) \simeq 1,  \label{S_asym1_1}  \\
&J_{1}(4t) \simeq 2t, \label{S_asym1_2}
\end{align}
which can be readily derived using the infinite series of $J_n(x)$. 
Substituting Eqs.~\eqref{S_asym1_1} and \eqref{S_asym1_2} into Eq.~\eqref{variance7} of the main text, we obtain, up to the leading order, 
\begin{align}
\sigma(t)^2 \simeq 2 t^2.
\end{align}
For the long-time dynamics ($t \gg 1$), we use the asymptotic formula \cite{math1} given by
\begin{align}
J_{n}(4t) = \frac{ \displaystyle \cos \left( 4t - \frac{\pi}{2} n - \frac{\pi}{4} \right)}{ \displaystyle \sqrt{2 \pi t}} \left( 1 - \frac{\Gamma(n+5/2)}{128 \Gamma(n-3/2)} t^{-2}  \right) - \frac{\displaystyle \sin \left( 4t - \frac{\pi}{2} n - \frac{\pi}{4} \right) }{ \displaystyle \sqrt{2 \pi} t^{3/2} }\frac{\Gamma(n+3/2)}{8 \Gamma(n-1/2)}  + \mathcal{O}(t^{-7/2}).
\label{S_asym2}
\end{align}
This expression leads to
\begin{align}
J_{0}(4t)^2           &\simeq \dfrac{1}{2\pi t} \cos \left( 4t - \dfrac{\pi}{4} \right)^2 + \dfrac{1}{64\pi t^2} \cos \left( 8t \right) + \dfrac{1}{2048\pi t^3} \sin \left( 4t - \dfrac{\pi}{4} \right)^2 - \dfrac{9}{2048 \pi t^3} \cos \left( 4t - \dfrac{\pi}{4} \right)^2, \\
J_{1}(4t)^2           &\simeq \dfrac{1}{2\pi t} \sin \left( 4t - \dfrac{\pi}{4} \right)^2 - \dfrac{3}{64\pi t^2} \cos \left( 8t \right)   + \dfrac{9}{2048\pi t^3} \cos \left( 4t - \dfrac{\pi}{4} \right)^2 + \dfrac{15}{2048 \pi t^3} \sin \left( 4t - \dfrac{\pi}{4} \right)^2, \\
J_{0}(4t)J_{1}(4t)  &\simeq  - \dfrac{1}{8 \pi t} \cos \left( 8t \right) + \dfrac{3}{64\pi t^2} \cos \left( 4t - \dfrac{\pi}{4} \right)^2 + \dfrac{1}{64 \pi t^2} \sin \left( 4t - \dfrac{\pi}{4} \right)^2.
\label{S_asym3}
\end{align}
Employing these expressions and Eq.~\eqref{variance7}, we derive
\begin{align}
\sigma(t)^2 \simeq \dfrac{2}{\pi} t - \dfrac{1}{64 \pi t} \left(  2 \sin (8t) + 1 \right).
\end{align}

\section{ Derivation of Eq.~(\ref{incomp_alt4}) } 
We shall derive the explicit expression of the two-point correlator with the incomplete alternating initial state $\hat{\rho}_{\rm alt}$ defined by Eq.~\eqref{incomp_alt1} of the main text.
We first define the two-point correlator $C_{m,n}(t)$ as
\begin{align}
C_{m,n}(t) \coloneqq {\rm Tr} \left[ e^{-{\rm i}\hat{H}t} \hat{\rho}_{\rm alt} e^{{\rm i}\hat{H}t} \hat{a}^{\dagger}_m \hat{a}_n \right].
\label{S_Cdef}
\end{align}
The straightforward calculation leads to the equation of motion for $ C_{m,n}(t) $, 
\begin{align}
{\rm i} \dfrac{d}{dt} C_{m,n}(t) = C_{m+1,n}(t) + C_{m-1,n}(t) - C_{m,n+1}(t) - C_{m,n-1}(t) . 
\label{S_Ceom}
\end{align}
To solve the differential equation, we expand the correlator via the discrete Fourier transformation, getting
\begin{align}
C_{m,n}(t) = \dfrac{1}{4L^2}\sum_{\alpha=-L}^{L-1} \sum_{\beta=-L}^{L-1} D_{\alpha,\beta}(t) e^{ {\rm i} \pi (n \beta - m \alpha)/L}, 
\label{S_dft}
\end{align}
where $D_{\alpha,\beta}(t) $ is the coefficient of the expansion. Substituting Eq.~\eqref{S_dft} into Eq.~\eqref{S_Ceom}, we readily derive
\begin{align}
{\rm i} \dfrac{d}{dt} D_{\alpha,\beta}(t) = \mathcal{E}_{\alpha, \beta} D_{\alpha,\beta}(t) 
\label{S_Deom}
\end{align}
with $ \mathcal{E}_{\alpha, \beta} \coloneqq 2 \cos (\pi \alpha/L) - 2 \cos (\pi \beta/L) $. 
The initial condition $D_{\alpha,\beta}(0)$ is calculated as
\begin{align}
D_{\alpha,\beta}(0) 
&=  \sum_{m=-L}^{L-1} \sum_{n=-L}^{L-1} C_{m,n}(0) e^{ {\rm i} \pi (-n \beta + m \alpha)/L} \\
&=  n_{\rm even} \sum_{m=-L/2}^{L/2-1} e^{ {\rm i} 2\pi m ( \alpha- \beta)/L} + n_{\rm odd} \sum_{m=-L/2}^{L/2-1} e^{ {\rm i} \pi (2m +1)( \alpha- \beta)/L}\\
&= L n_{\rm even} ( \delta_{\alpha,\beta} + \delta_{\alpha,\beta+L} + \delta_{\alpha,\beta-L} ) + L n_{\rm odd} ( \delta_{\alpha,\beta} - \delta_{\alpha,\beta+L} - \delta_{\alpha,\beta-L} ), 
\end{align}
where we use $-2L+1 \leq \alpha \pm \beta \leq 2L-1$ since the values of $\alpha$ and $\beta$ are restricted to the 1st Brillouin zone. As to the initial condition $C_{m,n}(0)$, by the definition of $\hat{\rho}_{\rm alt}$, we can obtain  
\begin{align}
C_{m,n}(0) =  
 \begin{dcases}
     n_{\rm even}     & ( n=m \wedge n~{\rm is ~even})     \\
     n_{\rm odd}     & ( n=m \wedge n~{\rm is ~odd}) \\
     0                 & ({\rm otherwise}), 
  \end{dcases}
\end{align}
where $n_{\rm even}$ and $n_{\rm odd}$ are averaged particle numbers at the even and odd sites for the initial density matrix $\hat{\rho}_{\rm alt}$, respectively. 
Solving the differential equation of Eq.~\eqref{S_Deom}, we get
\begin{align}
D_{\alpha,\beta}(t) &= D_{\alpha,\beta}(0) e^{ - {\rm i} \mathcal{E}_{\alpha, \beta} t}  \\
&= L n_{\rm T} \delta_{\alpha,\beta} + L n_{\rm D} \left( \delta_{\alpha, \beta+L} + \delta_{\alpha, \beta-L} \right)  e^{  {\rm i} 4t \cos (\pi \beta/L) }, 
\label{eq:D0}
\end{align}
where we define $n_{\rm T} \coloneqq n_{\rm even} + n_{\rm odd}$ and $n_{\rm D} \coloneqq n_{\rm even} - n_{\rm odd}$. 
We substitute Eq.~\eqref{eq:D0} into Eq.~\eqref{S_dft}, getting
\begin{align}
C_{m,n}(t) 
&= \dfrac{1}{4L^2}\sum_{\alpha=-L}^{L-1} \sum_{\beta=-L}^{L-1} D_{\alpha,\beta}(t) e^{ {\rm i} \pi (n \beta - m \alpha)/L}\\
&= \dfrac{1}{4L}\sum_{\alpha=-L}^{L-1} \sum_{\beta=-L}^{L-1} \left(   n_{\rm T} \delta_{\alpha,\beta} +  n_{\rm D} \left( \delta_{\alpha, \beta+L} + \delta_{\alpha, \beta-L} \right)  e^{  {\rm i} 4t \cos (\pi \beta/L) } \right) e^{ {\rm i} \pi (n \beta - m \alpha)/L}\\
&= \dfrac{n_{\rm T} }{4L}\sum_{\alpha=-L}^{L-1} e^{ {\rm i} \pi (n  - m ) \alpha/L}  
+ \dfrac{n_{\rm D} (-1)^n}{4L}\sum_{\alpha=0}^{L-1}   e^{  -{\rm i} 4t \cos (\pi \alpha/L) } e^{ {\rm i} \pi (n  - m )\alpha/L} + \dfrac{n_{\rm D} (-1)^n}{4L}\sum_{\alpha=-L}^{-1}   e^{  -{\rm i} 4t \cos (\pi \alpha/L) } e^{ {\rm i} \pi (n  - m )\alpha/L} \\
&= \dfrac{n_{\rm T} }{4L}\sum_{\alpha=-L}^{L-1} e^{ {\rm i} \pi (n  - m ) \alpha/L}  
+ \dfrac{n_{\rm D} (-1)^n}{4L}\sum_{\alpha=-L}^{L-1}   e^{  -{\rm i} 4t \cos (\pi \alpha/L) } e^{ {\rm i} \pi (n  - m )\alpha/L} \\
&= \dfrac{n_{\rm T} }{2} \delta_{m,n}  
+ \dfrac{n_{\rm D} (-1)^n}{4L}\sum_{\alpha=-L}^{L-1}   e^{  -{\rm i} 4t \cos (\pi \alpha/L) } e^{ {\rm i} \pi (n  - m )\alpha/L}, 
\end{align}
where we again use $-2L+1 \leq \alpha \pm \beta \leq 2L-1$. Finally, we take the thermodynamic limit ($L \rightarrow \infty$), obtaining 
\begin{align}
C_{m,n}(t) 
&= \dfrac{n_{\rm T} }{2} \delta_{m,n}  + \dfrac{n_{\rm D} (-1)^n}{4\pi} \int_{-\pi}^{\pi}  d \theta ~ e^{ {\rm i}  (n  - m ) \theta -{\rm i} 4 t \cos \theta } \\
&= \dfrac{n_{\rm T} }{2} \delta_{m,n}  + \dfrac{n_{\rm D} (-1)^m}{4\pi} \int_{0}^{2\pi}  d \theta ~ e^{ {\rm i}  (n  - m ) \theta +{\rm i} 4 t \cos \theta } \\
&= \dfrac{n_{\rm T} }{2} \delta_{m,n}  + \dfrac{n_{\rm D} }{2} {\rm i}^{n+m}  J_{n-m}(4t).
\label{eq:C0}
\end{align}
In the last line, we use the integral formula for the Bessel function of the first kind given by
\begin{align}
J_n(x) = \dfrac{1}{2\pi {\rm i}^n} \int_0^{2 \pi} d \theta ~ e^{ {\rm i} n \theta + {\rm i}x \cos \theta }. 
\end{align}

\section{Derivation of Eq.~(\ref{incomp_alt5}) } 
We shall derive Eq.~\eqref{incomp_alt5} of the main text using Eq.~\eqref{incomp_alt4} of the main text.
Note that the initial state $\hat{\rho}_{\rm alt}$ is the Gaussian state and thus we can use the Wick theorem. 
As a result, the variance is given by
\begin{align}
\sigma_{M}(t)^2 = \sum_{m=0}^{M-1} C_{m,m}(t) - \sum_{m=0}^{M-1} \sum_{n=0}^{M-1} |C_{m,n}(t)|^2.
\label{S_variance_exp1}
\end{align}
Following the almost same procedure as Sec.~\ref{sec:1}, we obtain
\begin{align}
\sum_{m=0}^{M-1} C_{m,m}(t) &= \dfrac{M}{2} (n_{\rm even} + n_{\rm odd}) + \dfrac{1}{2} (n_{\rm even} - n_{\rm odd}) J_0(4t) \sum_{m=0}^{M-1} (-1)^m, \\
\sum_{m=0}^{M-1} \sum_{n=0}^{M-1} |C_{m,n}(t)|^2 &= \dfrac{M}{4} (n_{\rm even} + n_{\rm odd})^2 + \dfrac{1}{2} (n_{\rm even}^2 - n_{\rm odd}^2) J_0(4t) \sum_{m=0}^{M-1} (-1)^m \\
&+ \dfrac{1}{4}(n_{\rm even} - n_{\rm odd})^2 \sum_{m=0}^{M-1} \sum_{n=0}^{M-1} J_{n-m}^2(4t)\\
&= \dfrac{M}{4} (n_{\rm even} + n_{\rm odd})^2 + \dfrac{1}{2} (n_{\rm even}^2 - n_{\rm odd}^2) J_0(4t) \sum_{m=0}^{M-1} (-1)^m \\
&+ \dfrac{M}{4}(n_{\rm even} - n_{\rm odd})^2 \left(  J_0(4t)^2 + 2\sum_{m=1}^{M-1} J_m(4t)^2 \right) - \dfrac{1}{2}(n_{\rm even} - n_{\rm odd})^2 \sum_{m=1}^{M-1} m J_m(4t)^2. 
\end{align}
Thus, the variance becomes
\begin{align}
\sigma_{M}(t)^2 &= \dfrac{M}{4} (n_{\rm even} + n_{\rm odd}) (2 - n_{\rm even} - n_{\rm odd}) + \dfrac{1}{2} ( n_{\rm even} - n_{\rm odd} - n_{\rm even}^2 + n_{\rm odd}^2 ) J_0(4t) \sum_{m=0}^{M-1} (-1)^m \\
& - \dfrac{M}{4}(n_{\rm even} - n_{\rm odd})^2 \left(  J_0(4t)^2 + 2\sum_{m=1}^{M-1} J_m(4t)^2 \right) + \dfrac{1}{2}(n_{\rm even} - n_{\rm odd})^2 \sum_{m=1}^{M-1} m J_m(4t)^2.
\end{align}
Next, we define $\delta \sigma_M(t)^2$ as
\begin{align}
\delta \sigma_M(t)^2 \coloneqq \dfrac{M}{2} \bigl(  n_{\rm even}( 1- n_{\rm even} ) + n_{\rm odd} (1 - n_{\rm odd}) \bigl) + \dfrac{1}{2} J_0(4t) \bigl(  n_{\rm even}( 1- n_{\rm even} ) - n_{\rm odd} (1 - n_{\rm odd}) \bigl) \sum_{m=0}^{M-1} (-1)^m.
\end{align}
As a result, we can obtain 
\begin{align}
\sigma_{M}(t)^2  - \delta \sigma_M(t)^2 = (n_{\rm even} - n_{\rm odd})^2 \left[ \dfrac{M}{4} \left( 1 - J_0(4t)^2 - 2 \sum_{m=1}^{M-1}J_m(4t)^2 \right) + \dfrac{1}{2} \sum_{m=1}^{M-1} m J_m(4t)^2 \right].
\end{align}
Finally, applying Eqs.~\eqref{S_lim1} and \eqref{S_lim2} into the above, we derive Eq.~\eqref{incomp_alt5} of the main text, namely
\begin{align}
\lim_{M \rightarrow \infty} \bigl( \sigma_{M}(t)^2  - \delta \sigma_M(t)^2  \bigl) = (n_{\rm even} - n_{\rm odd})^2 \bigl[ 4t^2 \left(  J_0(4t)^2 +  J_1(4t)^2 \right) - t  J_0(4t) J_1(4t) \bigl].
\end{align}

\section{Numerical perturbative calculation for the disorder potential }
We shall explain how to implement the numerical perturbative calculation for the disorder potential. 
In the main text, we define the random potential $V_j$, which is independently sampled from the uniform distribution with the range $[-\Delta, \Delta]$. 
Here, $\Delta \geq 0$ denotes the strength of the randomness. This definition is inconvenient to proceed perturbative calculation for the disorder potential, so thus we introduce a new random variable $v_n$, which is independently sampled from the uniform distribution with the range $[-1,1]$. Then, the Hamiltonian $\hat{H}'$ of Eq.~\eqref{Hamil_exp} in the main text is written as
\begin{align} 
\hat{H}' = -\sum_{j = -L }^{L-1} \left( \hat{a}_{j+1}^{\dagger} \hat{a}_j + \hat{a}_{j}^{\dagger} \hat{a}_{j+1}   \right) +  \Delta \sum_{j = -L }^{L-1} v_j \hat{n}_{j}.
\label{S_Hamil_exp}
\end{align}
Using this notation, we can derive the equation of motion for the two-point correlator $C_{m,n}(t)$
\begin{align}
{\rm i} \dfrac{d}{dt} C_{m,n}(t) = C_{m+1,n}(t) + C_{m-1,n}(t) - C_{m,n+1}(t) - C_{m,n-1}(t) + \Delta ( v_n - v_m ) C_{m,n}(t).
\label{S_CeomV}
\end{align}
In what follows, we regard $\Delta$ as a small parameter and construct the perturbative equation of motion for the two-point correlator.  

\begin{figure}[t]
\begin{center}
\includegraphics[keepaspectratio, width=14.0cm]{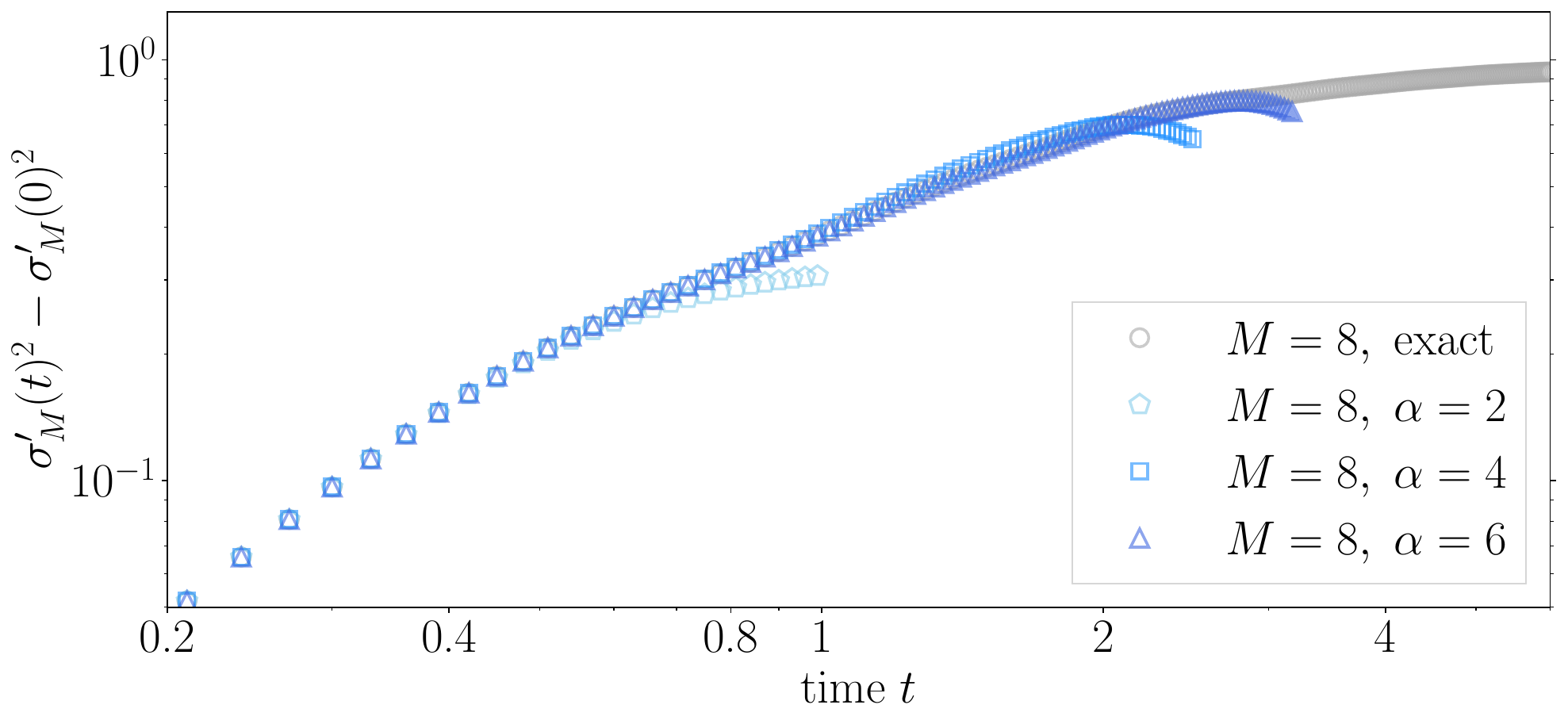}
\caption{
Numerical results of the variance for the perturbative calculations with $\Delta=1$. The parameters are the same as those used in Fig.~3 of the main text. 
The pentagon, square, and triangle markers denote the data of the variance $\sigma'_{M}(t)^2$ for the $\alpha=2, 4,$ and $6$th-order perturbative calculations, respectively, and the circle marker does the exact numerical data. 
Here, the exact result is numerically obtained without using the perturbation calculation.
}  
\label{Sfig2} 
\end{center}
\end{figure}

We expand the two-point correlator with the parameter $\Delta$ as 
\begin{align}
C_{m,n}(t) = C_{m,n}^{(0)}(t) + \Delta  C_{m,n}^{(1)}(t) + \Delta^2  C_{m,n}^{(2)}(t) + ...~~.
\label{S_Cexp1}
\end{align}
Substituting Eq.~\eqref{S_Cexp1} into Eq.~\eqref{S_CeomV}, we obtain the $\alpha$th-order equation of motion as  
\begin{align}
{\rm i} \dfrac{d}{dt} C_{m,n}^{(\alpha)}(t) = C_{m+1,n}^{(\alpha)}(t) + C_{m-1,n}^{(\alpha)}(t) - C_{m,n+1}^{(\alpha)}(t) - C_{m,n-1}^{(\alpha)}(t) + ( v_n - v_m ) C_{m,n}^{(\alpha-1)}(t), 
\label{S_pCeomV}
\end{align}
where we define $C_{m,n}^{-1}(t) \coloneqq 0$. 
When you calculate the variance up to the $\alpha$th-order, we numerically solve Eq.~\eqref{S_pCeomV} for $C_{m,n}^{(0)}(t), C_{m,n}^{(1)}(t),..., C_{m,n}^{(\alpha)}(t)$, by the 4th-order Runge-Kutta method, obtaining the $\alpha$th-order two-point correlator as
\begin{align}
\bar{C}_{m,n}(t) \coloneqq \sum_{\beta=0}^{\alpha} \Delta^{\beta}  C_{m,n}^{(\beta)}(t).
\label{S_Cexp2}
\end{align}
Then, the variance $\sigma_M'(t)^2$ defined in the main text can be computed using 
\begin{align}
&\braket{\hat{N}_M'}      =  \sum_{m=-M/2}^{M/2-1}  \bar{C}_{m,m}(t), \\
&\braket{ \hat{N}_M'^2} = \braket{\hat{N}_M'} + \braket{\hat{N}_M'}^2 - \sum_{m=-M/2}^{M/2-1} \sum_{n=-M/2}^{M/2-1}  |\bar{C}_{m,n}(t)|^2.
\end{align}

Figure~\ref{Sfig2} displays the numerical results of the variance for the 2nd-, 4th-, and 6th-order perturbative calculations with $\Delta=1$. One can see that the agreement of the perturbative calculations with the exact one becomes better as the order $\alpha$ increases.

\clearpage
\section{Effect of the superposition of the initial state in the noninteracting fermions} 
We numerically study the variance growth in the noninteracting fermions using a generalized incomplete initial state, which can include the superpositions between different sites. The setup is the same as the one used in Fig.~\ref{fig3} of the main text except for the initial state. We here use the generalized initial density matrix as an initial state, which is defined by
\begin{align}
\hat{\rho}_{\rm two{\text-}site} &\coloneqq \dfrac{1}{\Xi} \exp \left( -\beta \hat{H}_{\rm two{\text-}site} + \beta \mu \hat{N}_{\rm tot} \right).
\label{S_g_incomp1}
\end{align}
Here, the Hamiltonian $\hat{H}_{\rm two{\text-}site} $ and the total number operator $\hat{N}_{\rm tot}$ are given by
\begin{align}
\hat{H}_{\rm two{\text-}site} &\coloneqq - J \sum_{j=-L/2}^{L/2-1} \left( \hat{a}^{\dagger}_{2j+1} \hat{a}_{2j} + \hat{a}^{\dagger}_{2j} \hat{a}_{2j+1} \right) + \sum_{j=-L/2}^{L/2-1} \left( V_{\rm odd} \hat{a}^{\dagger}_{2j+1} \hat{a}_{2j+1} + V_{\rm even} \hat{a}^{\dagger}_{2j} \hat{a}_{2j} \right), \label{S_g_incomp2} \\
\hat{N}_{\rm tot} &= \sum_{j=-L}^{L-1} \hat{a}^{\dagger}_{j} \hat{a}_{j} \label{S_g_incomp3}
\end{align}
with the two-site hopping parameter $J$, the even-site potential $V_{\rm even}$, and the odd-site potential $V_{\rm odd}$. In the following, we set $\beta=1$ without loss of generality. 
For a given $J$, the values of $V_{\rm even}$, $V_{\rm odd}$, and $\mu$ are numerically determined such that $n_{\rm even}$ and $n_{\rm odd}$ satisfy the following conditions:
\begin{align}
I = \dfrac{n_{\rm odd}- n_{\rm even}}{n_{\rm odd} + n_{\rm even}} = 0.91, ~~~~ \nu = \dfrac{n_{\rm odd} + n_{\rm even}}{2} = 0.44.
\end{align}
For $J=0$, this generalized initial state is identical to the initial density matrix used in the main text. 
Under this situation, we can study the effect of the initial superposition on the variance growth in the noninteracting fermions. 

\begin{figure}[b]
\begin{center}
\includegraphics[keepaspectratio, width=14.0cm]{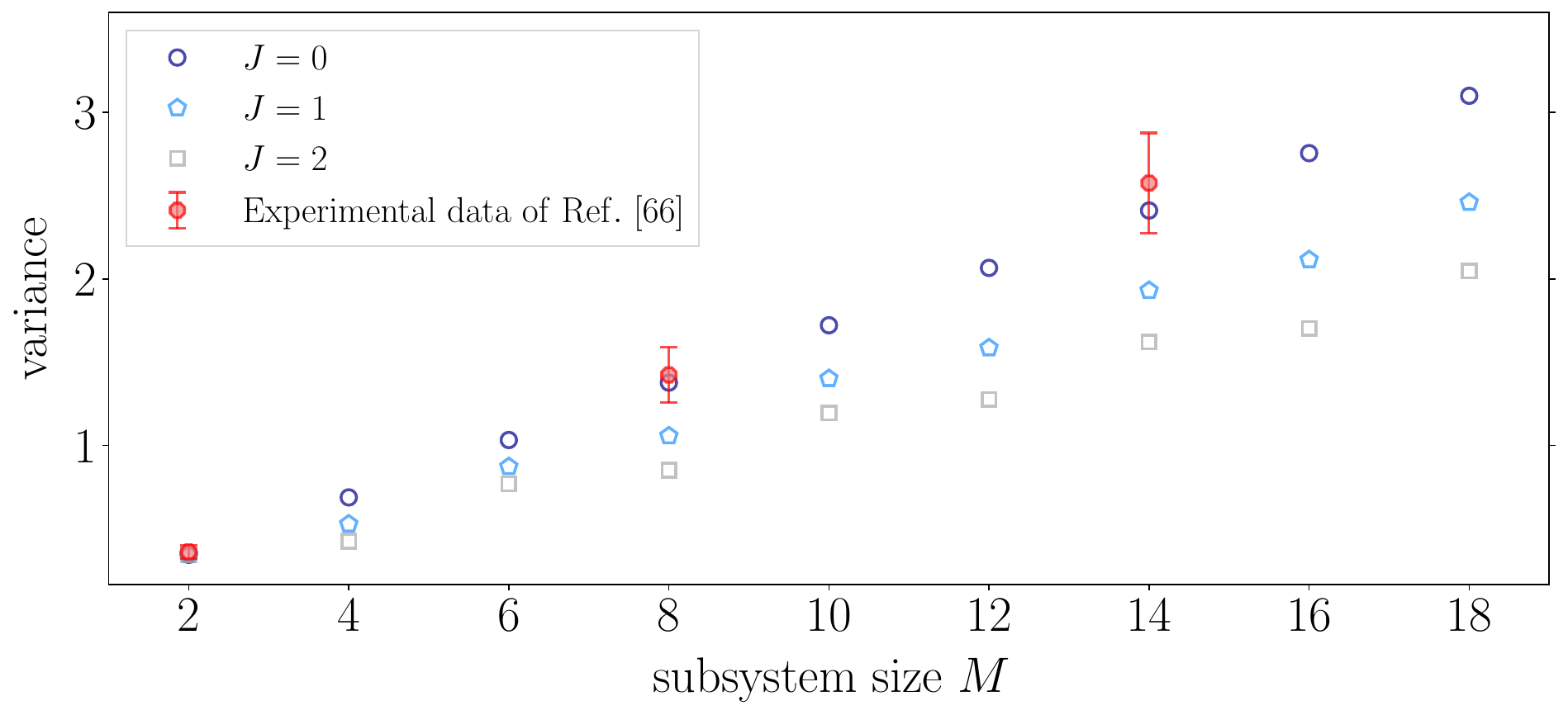}
\caption{
Numerical results for the dependence of the initial variance with the generalized initial state of Eq.~\eqref{S_g_incomp1} on the subsystem size $M$ and the initial hopping parameter $J$.  The circle, pentagon, and square markers represent the initial variance obtained numerically with $ J=0, 1, $ and $2$, respectively. The red hexagon markers with the bar is the experimental data for Fig.~S8 of Ref.~\cite{hydro}. For the comparison with the experiment, we plot the variance $2\sigma'_M(t)^2 $ for the numerical results and $\sigma_{\rm exp}(t)^2$ for the experimental result because the experiment studies two-ladder lattice systems and thus the observed variance is twice as large as ours. This graph clearly shows that the deviations between the experimental and numerical results become large when the initial hopping parameter $J$ increases. 
}  
\label{Sfig3} 
\end{center}
\end{figure}

\begin{figure}[t]
\begin{center}
\includegraphics[keepaspectratio, width=14.0cm]{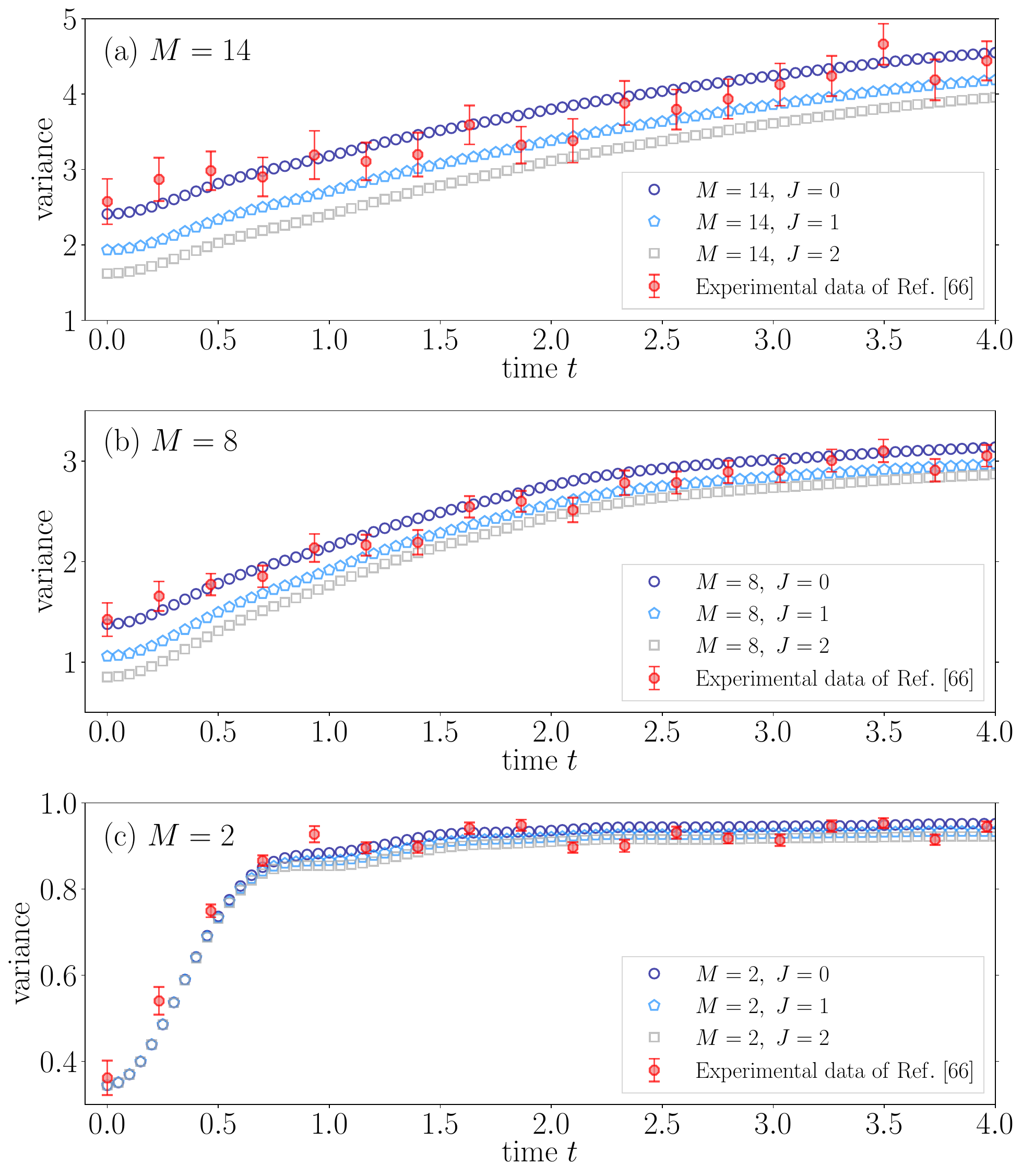}
\caption{
Numerical results for the variance growth in the noninteracting fermions with the generalized initial state of Eq.~\eqref{S_g_incomp1}.
The figures display the time evolution of the variance growth for (a) $M=14$, (b) $M=8$, and (c) $M=2$. 
For the comparison with the experiment, we plot the variance $2\sigma'_M(t)^2 $ for the numerical results and $\sigma_{\rm exp}(t)^2$ for the experimental results because the experiment studies two-ladder lattice systems and thus the observed variance is twice as large as ours.
The numerical results with $ J = 0$ exhibit better agreement with the experimental data for Fig.~S8 of Ref.~\cite{hydro} compared to the numerical data with the finite hopping parameter $J \neq 0$. 
}  
\label{Sfig4} 
\end{center}
\end{figure}

Figure~\ref{Sfig3} displays the dependence of the initial variance on the subsystem size $M$ and the hopping parameter $J$, clearly showing that the experimental date is very close to the numerical data with $J=0$. We also plot the time evolution of the variance growth with the generalized initial state of Eq.~\eqref{S_g_incomp1} in Fig.~\ref{Sfig4}.
Then, we find the numerical results with $J=0$ exhibit better agreement with the experimental data of Ref.~\cite{hydro}. 
All these numerical findings suggest that the initial superpositions of the experiment is strongly suppressed, as pointed out in the experimental paper of Ref.~\cite{hydro}.  

\section{Effect of the random potential on the interacting fermions} 
We numerically study how the random potentials affect the variance growth in the interacting fermions. 
The setup in this section is the same as the one used in Fig.~\ref{fig2} of the main text, but we use the Hamiltonian $\hat{H}_{\rm random}$ with the random potential, which is defined by 
\begin{align} 
\hat{H}_{\rm random} \coloneqq -\sum_{j = -L }^{L-1} \left( \hat{a}_{j+1}^{\dagger} \hat{a}_j + \hat{a}_{j}^{\dagger} \hat{a}_{j+1}   \right) +  \sum_{j = -L }^{L-1} V_j \hat{n}_{j} + U \sum_{j = -L }^{L-1} \hat{n}_{j+1} \hat{n}_{j}, 
\label{S_Hamil} 
\end{align}
where the random potential $V_j$ is independently sampled from the uniform distribution with the range $[-\Delta,\Delta]$. Here, $\Delta$ is the real positive parameter. 
We solve the Schrödinger equation with Eq.~\eqref{S_Hamil} via the TEBD method, numerically investigating the effect of the random potential on the variance growth. 
In what follows, the time evolution of the variance is computed by using 200 samples for the ensemble average of the random potential.  

\begin{figure}[t]
\begin{center}
\includegraphics[keepaspectratio, width=14.0cm]{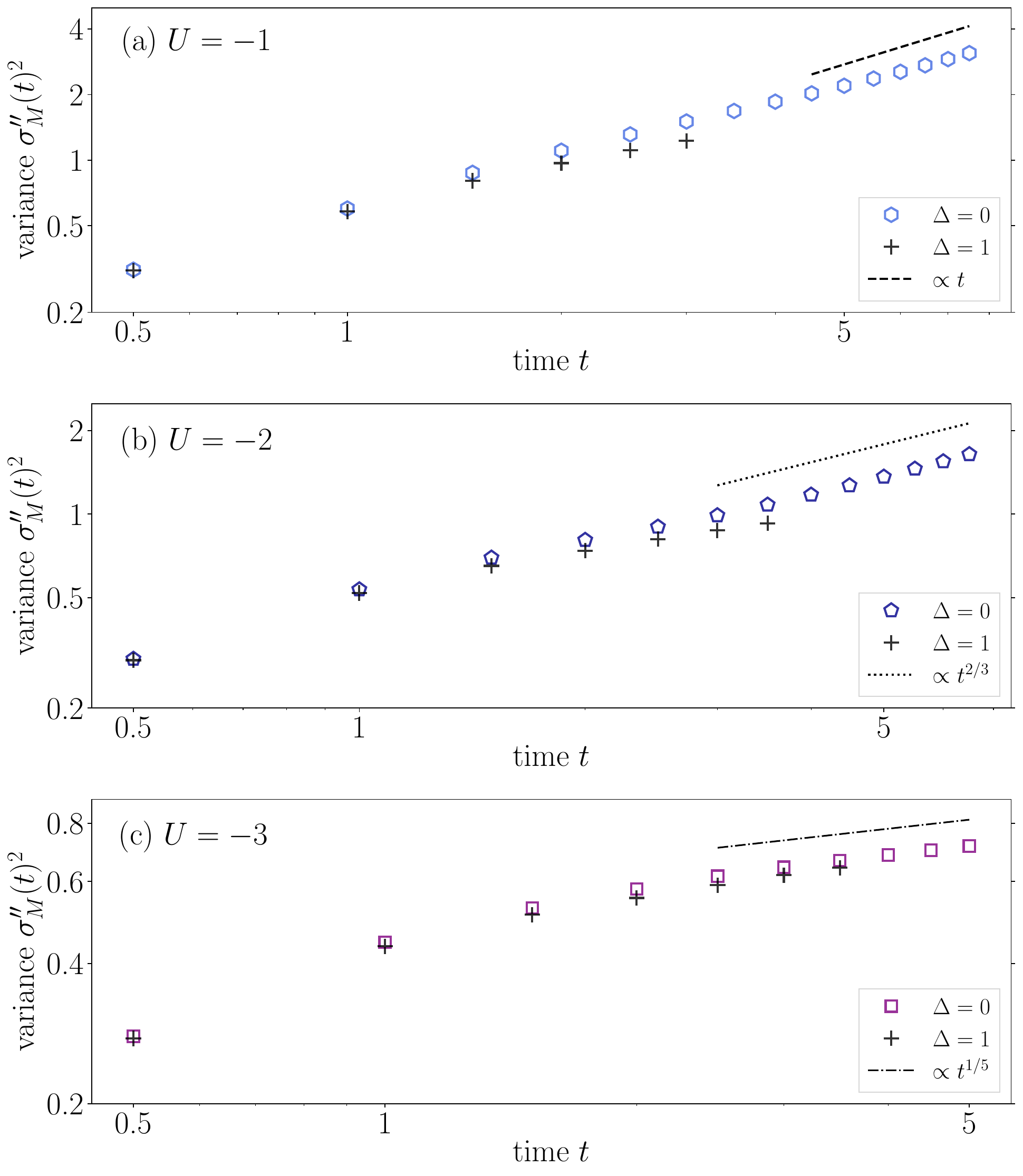}
\caption{ 
Numerical results of the variance growth in the interacting fermions with the random potential for (a) $U=-1$, (b) $U=-2$, and (c) $U=-3$. 
In each graph, we put the data with the strength $\Delta=0$ and $1$ of the random potential to make visible the disorder effect clearly. 
The numerical data with $\Delta=0$ is the same as the data of Fig.~\ref{fig2} of the main text. 
}  
\label{Sfig5} 
\end{center}
\end{figure}

Figure~\ref{Sfig5} shows the numerical results for the variance growth with $\Delta=0$ and $1$. Our numerical simulations for the interaction parameter $U = -1$ and $-2$ shows that the disorder effect becomes large as time goes by. Thus, in these cases we expect that the power-law-like growth seen in the interacting fermions with $\Delta = 0$ will be strongly disturbed in the long-time dynamics. On the other hand, the disorder effect for $U=-3$ is not large in the time region we can numerically access.

\section{Dependence on the bond dimension and the time resolution in the numerical calculations}
\begin{figure}[t]
\begin{center}
\includegraphics[keepaspectratio, width=18.0cm]{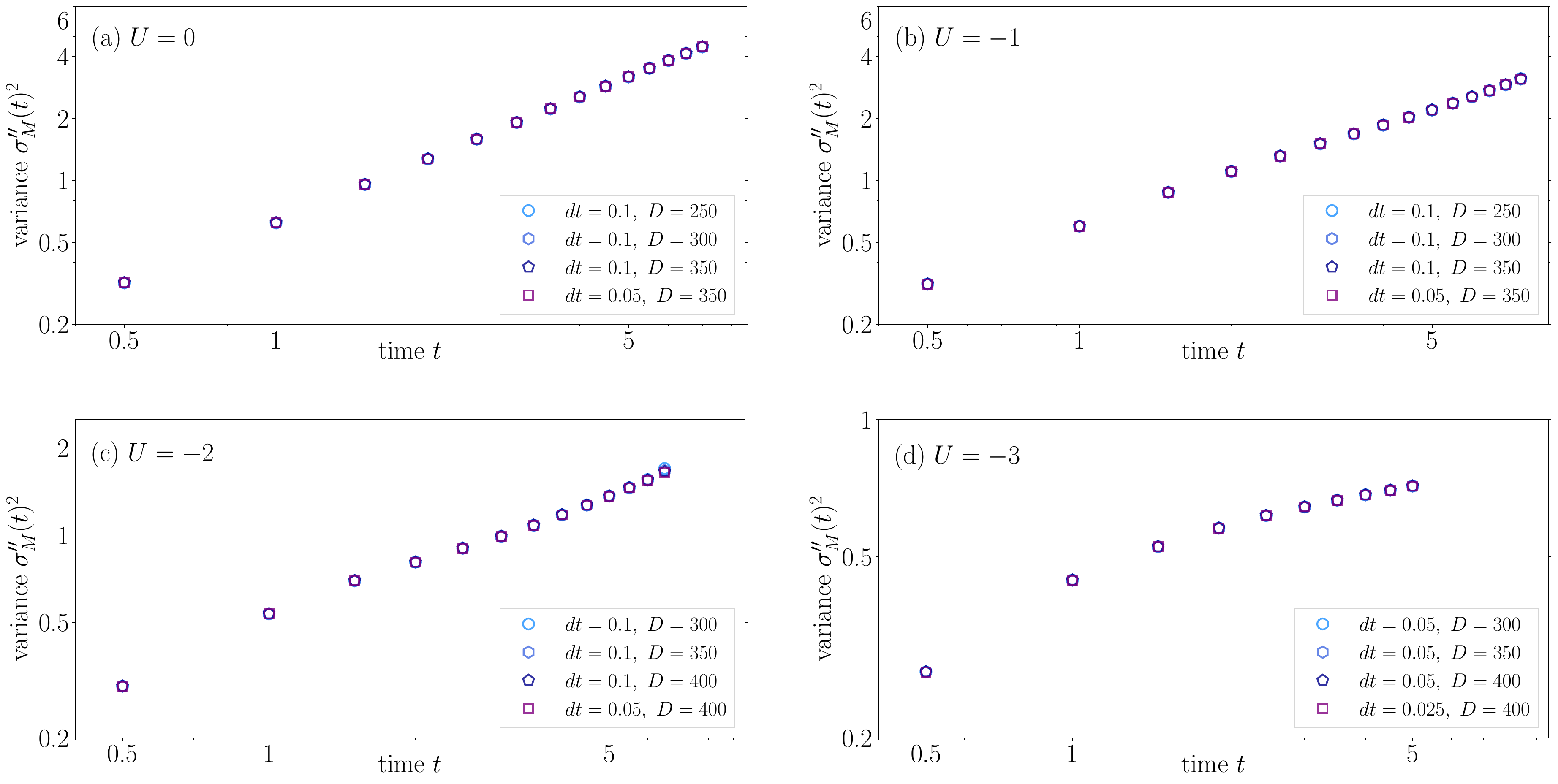}
\caption{ 
Dependence on the bond dimension $D$ and the time resolution $dt$ for the variance growth in the interacting fermions with (a) $U=0$, (b) $U=-1$, (c) $U=-2$, and (d) $U=-3$. 
}  
\label{Sfig6} 
\end{center}
\end{figure}

We describe the details of our numerical simulation based on the TEBD method. 
The notations used in this section are the same as those used in Fig.~\ref{fig2} of the main text.
In the TEBD method, we usually express a quantum state $\ket{\psi}$ as
\begin{align} 
\ket{\psi} = \sum_{\alpha_1=1}^{D_1} \sum_{n_1=0}^{1} \cdots \sum_{\alpha_1=1}^{D_{L-1}} \sum_{n_{L-1}=0}^{1} \sum_{n_{L}=0}^{1} \Gamma_{\alpha_1}^{[1], n_1} \lambda_{\alpha_1}^{[1]} \Gamma_{\alpha_1,\alpha_2}^{[2], n_2} \lambda_{\alpha_2}^{[2]} \cdots \Gamma_{\alpha_{L-2},\alpha_{L-1}}^{[L-1], n_{L-1}} \lambda_{\alpha_{L-1}}^{[L-1]} \Gamma_{\alpha_{L-1}}^{[L],n_L}
\label{S_MPS1} 
\end{align}
with the real variable $\lambda$ and the complex variable $\Gamma$, where their subscripts are omitted for brevity. The parameter $D_j~({j \in \{1,..., L-1 \}})$ is referred to as a bond dimension and plays the role of a cutoff parameter in the numerical simulation. Here, we emphasize that our model conserves the total particle number but the expression of Eq.~\eqref{S_MPS1} does not incorporate the merit of such conservation law. In order to utilize the conservation law, it is helpful to consider the following expression instead of Eq.~\eqref{S_MPS1}:
\begin{align} 
\ket{\psi} = \sum_{q_1} \sum_{\alpha_1=1}^{D_1(q_1)} \sum_{n_1=0}^{1} \cdots \sum_{q_{L-1}} \sum_{\alpha_1=1}^{D_{L-1}(q_{L-1})} \sum_{n_{L-1}=0}^{1} \sum_{q_{L}} \sum_{n_{L}=0}^{1} \Gamma_{\alpha_1}^{[1], n_1, q_1} \lambda_{\alpha_1}^{[1], q_1} \Gamma_{\alpha_1,\alpha_2}^{[2], n_2, q_2} \lambda_{\alpha_2}^{[2], q_2} \cdots \Gamma_{\alpha_{L-2},\alpha_{L-1}}^{[L-1], n_{L-1}, q_{L-1}} \lambda_{\alpha_{L-1}}^{[L-1], q_{L-1}} \Gamma_{\alpha_{L-1}}^{[L],n_L, q_L}, 
\label{S_MPS2} 
\end{align}
where we introduce the new integer $q_j~(j \in \{1,..., L \})$ to reduce the computational costs. The critical point of this expression is that the bond dimension $D_{j} (q_j)$ depends on $q_j$. In our numerical calculation, we introduce the maximal bond dimension $D$ such that all the bond dimensions $D_{j} (q_j)$ are smaller than $D$. Note that we need a larger bond dimension when we do not consider the particle-number conservation in the numerical calculation. As to the time discretization, we use the 2nd-order checkerboard discretization based on the Suzuki-Trotter decomposition with the time resolution $dt$. 

Figure~\ref{Sfig6} displays the time evolution of the variance corresponding to Fig.~\ref{fig2} of the main text. Then, one can see that the numerical data show almost the same behavior irrespective of the values of $D$ and $dt$. Hence, our numerical results are well convergent. In Fig.~\ref{fig2} of the main text, we use the numerical data with the values of $D$ and $dt$ summarized in Table~\ref{table:TEBD}.

\begin{table}[hbtp]
  \caption{ Values of $D$ and $dt$ used for the numerical data of Fig.~\ref{fig2} of the main text }
  \label{table:TEBD}
  \centering
  \begin{tabular}{ccc}
    \hline
    Interaction parameter $U$~~~~ & Time resolution $dt$~~~~  &  Bond dimension $D$~~~~  \\
    \hline \hline
    0   & 0.05  & 350 \\
    -1  & 0.05  & 350 \\
    -2  & 0.05  & 400 \\
    -3  & 0.05  & 400 \\
    \hline
  \end{tabular}
\end{table}

\end{document}